\documentclass[iop]{emulateapj}
\providecommand{\e}[1]{\ensuremath{\times 10^{#1}}}
\usepackage{color}
\usepackage{soul}
\usepackage{ulem}
\usepackage{amssymb,amsmath}
\begin{document}

\title{Tackling Radio Polarization of Energetic Pulsars}
\author{H. A. Craig}
\affiliation{Stanford University}

\begin{abstract}
The traditional, geometrical rotating vector model (RVM) has proved
particularly poor at capturing the polarization sweeps of the young energetic
and millisecond pulsars detected by \textit{Fermi}. We augment this model by including
finite altitude effects using a swept back vacuum dipole geometry. By further
including the effects of orthogonal mode jumps, multiple emission altitudes, open
zone growth via y-point lowering, and interstellar scattering, we show that a wide
range of departures from RVM can be modeled well while retaining a geometrical
picture. We illustrate these effects by fitting six \textit{Fermi}-detected pulsars 
(J0023$+$0923, J1024$-$0719, J1744$-$1134, J1057$-$5226, J1420$-$6048, and J2124$-$3358) and 
we describe how such modeling can improve our understanding of their emission geometry.
\end{abstract}
\section{Introduction}
\label{sec:intro}

For pulsars emitting in the radio, the conventional assumption 
is that the electric vector position angle 
follows the projection onto the plane of the sky from 
the magnetic field line at the emission point.  
Polarization position angle curves (position angle versus pulsar phase), which we see
as the pulsar sweeps past our field of view,
are very closely related to the orientation of the magnetic field lines.
Analysis of radio polarization is a powerful tool for understanding the geometry of pulsars.
For example, polarization contains information about 
the phase of closest approach of the surface dipole axis.  
Additionally, polarization has traditionally been used to 
place strong constraints on the impact angle, the angle 
between the magnetic pole of the pulsar, and the viewing direction.
Modeling polarization should also give estimates for the geometric 
parameters $\alpha$, the angle of magnetic axes, and $\zeta$, the viewing angle.

Naively, such projections should result in smooth polarization curves 
versus the pulsar period phase, particularly when adopting a point dipole model.
We argue that zero altitude models are not appropriate for certain pulsars. 
In stark contrast, a relatively recent paper \cite{yan2011polarization} exhibits 
the multitude of shapes that occur in millisecond pulsar polarization. 
More subtly, both polarization angle sweeps originating from zero altitude and 
polarization angle sweeps originating from a single, finite altitude can 
differ significantly in shape, although both appear smooth.  
Emission from finite altitude is a consideration for both millisecond pulsars and young pulsars.
\cite{karastergiou2007empirical} give the emission altitude of young pulsars
as $950$--$1000$ km. This emission altitude is then 
$0.02 \times 1000 / P_{\rm{ms}}$ $R_{\rm{LC}}$ in terms of the light cylinder radius.
The light cylinder radius, $R_{\rm{LC}}$, is the distance from the center of 
the neutron star at which co-rotating particles would be traveling
at the speed of light.
Since  $P_{\rm{ms}} < 100$ ms for young pulsars, their emission altitude would
be $> 0.2 R_{\rm{LC}}$, a significant fraction of the light cylinder. 
The neutron star radius in terms of light cylinder is $0.02 \times 10 / P_{\rm{ms}}$ $R_{\rm{LC}}$ 
for a neutron star radius of $10$ km.  For millisecond pulsars
with $P_{\rm{ms}} < 5$ ms, emission must come from $>0.04 R_{\rm{LC}}$, 
which is also a significant fraction of the light cylinder.

Precise modeling of millisecond pulsar and young pulsar radio 
polarization is of particular interest now because of the 
growth of $\gamma$-ray data from the \textit{Fermi Gamma-Ray Space Telescope}. 
These energetic pulsars make up the $\gamma$-ray pulsar population.
Thus, the understanding of $\gamma$-ray models can potentially 
benefit from radio polarization modeling because 
of the constraints on geometry that polarization often provides.
All pulsars considered in this paper are \textit{Fermi}-detected pulsars.

In essence, the present paper is an extension of the \cite{karastergiou2009complex}
paper in which the author shows one can produce theoretical 
polarization curves similar to those observed using 
orthogonal mode jumps and interstellar scattering. 
Here, we also allow for emission from finite altitude (numerically calculated).  
Although analytically calculated modifications exist for small altitude emission, 
such calculations contain estimates that break down at altitudes $\sim 0.1R_{\rm{LC}}$.
Our model also allows for multiple altitudes of emission.  Differences in 
altitude can explain non-$90^\circ$ position angle jumps seen 
particularly in millisecond pulsar polarization data.
Another major difference between \cite{karastergiou2009complex} and the present paper is that we 
seek to quantitatively fit the model to the data 
resulting in parameters with error bars and $\chi^2$ estimates. 
In contrast, \cite{karastergiou2009complex} was satisfied with producing polarization 
sweeps that appeared qualitatively similar to the data.
Further, using the \textit{F}-test, we compare the $\chi^2$ of the simplistic point dipole model 
and our more complex model to statistically quantify
whether the modifications are significant.
This paper is a methods paper that chooses pulsars that can 
clearly illustrate the strengths of this model; we do not tackle a large sample.

In Section~\ref{sec:RVM} of this paper, we discuss the rotating
vector model (RVM) and how the discrepancies between
data and the model demand a reevaluation of RVM.
In Section~\ref{sec:adding}, we describe the constituents of the
model in detail.  In Section~\ref{sec:fit}, we describe
the nuances of fitting the model.  Section~\ref{sec:ypt}
focuses on a parameter $\rho_{\rm{ypt}}$ which we define and use heavily
in this paper and which is a measure of the extent of 
the effective open zone required by phase of emission.  We apply the model
to data in Section~\ref{sec:app}. Table~\ref{tb:pulsarParam} gives
property parameters to the pulsars analysed.

\section{Rotating Vector Model and Beyond}
\label{sec:RVM}
We will start by discussing the analytic models used for 
radio position angle polarization and their shortcomings 
and then transition into the numerical model used for this paper.
The model predominately used for radio position angle polarization
is the RVM which
was formulated by \cite{radhakrishnan1969magnetic}.  The RVM is simple
and states that pulsars are point dipoles with emission
from the surface of the neutron star.  The
analytic RVM formula for polarization angles ($\psi$) is
\begin{equation}\label{eq:RVM}
\psi=\arctan\left[\frac{-\sin(\alpha)\sin(\phi+\Delta\phi)}
{\sin(\zeta) \cos(\alpha) -\cos(\zeta) \sin(\alpha) \cos(\phi+\Delta\phi)}\right]+\Delta\psi,
\end{equation}
where the inclination angle between the rotation
axis and magnetic axis is $\alpha$, the viewing angle is
$\zeta$, and the pulse phase is $\phi$. Measures of horizontal and vertical offset
are contained in $\Delta\psi$ and $\Delta\phi$.  These are the absolute phase and position angle on the 
sky of the magnetic axis.

        \begin{table}[ht]
	\small
        \caption{Property Parameters of the Pulsars}
        \begin{center}
        {
        \begin{tabular}{lcccc}
        \hline & \\[-1em]\hline

        Name            &Period     &$R_{\rm{LC}}/R_{\rm{NS}}$     &$f$       &DM           \\
			&(ms)	    &			 &(GHz)	    &(cm$^{-3}$ pc) \\
        [.3em]\hline 
        J0023$+$0923      &3.05           &12.5                    &1.649          &14.326                 
\\ \hline
        J1024$-$0719      &5.162          &20.0                    &1.369          &6.49                   
\\ \hline
        J1057$-$5226      &197.11         &1000                   &1.5            &30.1                   
\\ \hline
        J1744$-$1134      &4.075          &16.7                    &1.369          &3.14                   
\\ \hline
        J1420$-$6048      &68             &250                   &1.5 and 3      &360                    
\\ \hline
        J2124$-$3358      &4.931          &20.0                    &1.369          &4.60                   
\\ \hline

        \end{tabular}}
        \label{tb:pulsarParam}
        \end{center}
        \end{table}


Despite its simplicity, the RVM has been applied to numerous pulsars with great
success (i.e., \citealp{lyne1988shape}; \citealp{phillips1990magnetic}; \citealp{everett2001emission}).  
These pulsars are generally old, spun-down pulsars
with long periods and low altitudes of emission.
\cite{blaskiewicz1991relativistic} (the Blaskiewicz, Cordes, \& Wassermann, or BCW model) 
modified the RVM to include
finite altitude and found that the point of fastest change
in the polarization position angle sweep will shift back in phase
due to sweep-back effects on the magnetic field lines, while the intensity
profile shifts forward in phase due to co-rotation of the particles
in the pulsar magnetosphere.  Therefore, by fitting the position angle
data to the RVM and measuring this shift, one can estimate altitude.

The BCW formula with altitude ($r$) dependence measured in $R_{\rm{LC}}$ is given by

\begin{equation}\label{eq:BCWPhi}
\psi=\arctan\left[\frac{-\sin(\alpha)\sin(\phi-2r)}
{\sin(\zeta) \cos(\alpha) -\cos(\zeta) \sin(\alpha) \cos(\phi-2r)}\right]+\Delta\psi
\end{equation}
\citep{dyks2008altitude}.

\begin{figure}[t!!]
\vskip .40\textheight
\includegraphics{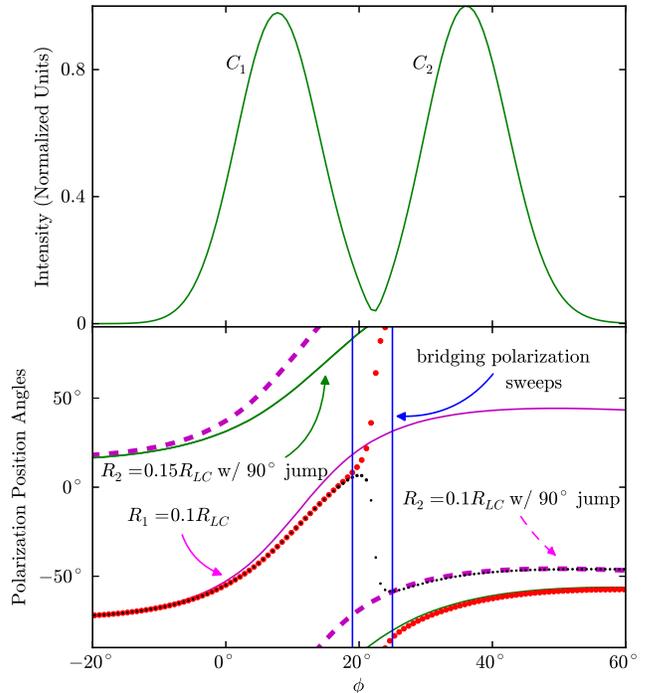}
\begin{center}
\caption{\label{fig:PlotJexampleintPA}
Plot of model pulsar intensity and polarization sweep showing the effects of single and 
multiple altitudes with a small scattering constant.
The model parameters are $\alpha=145^\circ$, $\zeta=140^\circ$, $P=5$ ms, and $\tau=0.03$ ms.
The black points are polarization position angles for a model with $R_1=0.1R_{\rm{LC}}$ for
polarization associated with intensity component $C_1$ 
and $R_2=0.1R_{\rm{LC}}$ plus an orthogonal mode jump for 
polarization associated with intensity component $C_2$.
The red points are polarization position angles for a model with $R_1=0.1R_{\rm{LC}}$ for 
polarization associated with intensity component $C_1$ 
and $R_2=0.15R_{\rm{LC}}$ plus an orthogonal mode jump for 
polarization associated with intensity component $C_2$.
Each model polarization sweep for a given component is
weighted using the Gaussian intensity profile.
Subtle changes in altitude can create drastic changes
in the direction of the bridging polarization in the phase of orthogonal mode jump.
The two Gaussian components of model intensity are equal in amplitude
but interstellar scattering effects make the first Gaussian component in 
phase ($C_1$) slightly lower in amplitude compared with the second
Gaussian component in phase ($C_2$). A phase of zero is the point
of closest encounter to the magnetic axis in the model.
}
\end{center}
\vskip -.3truecm
\end{figure}

\begin{figure*}[t!!]
\vskip .24\textheight
\includegraphics{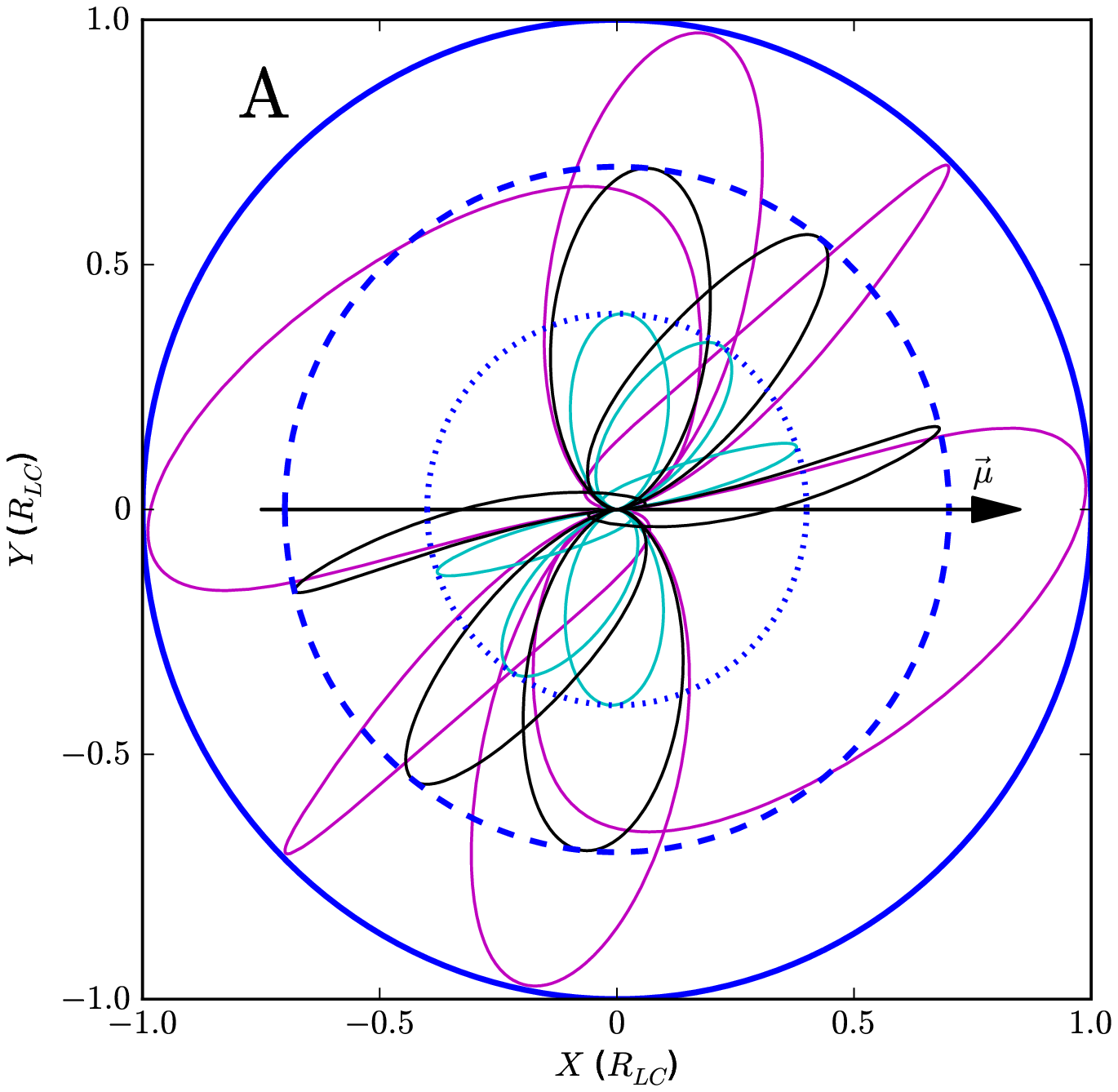}
\includegraphics{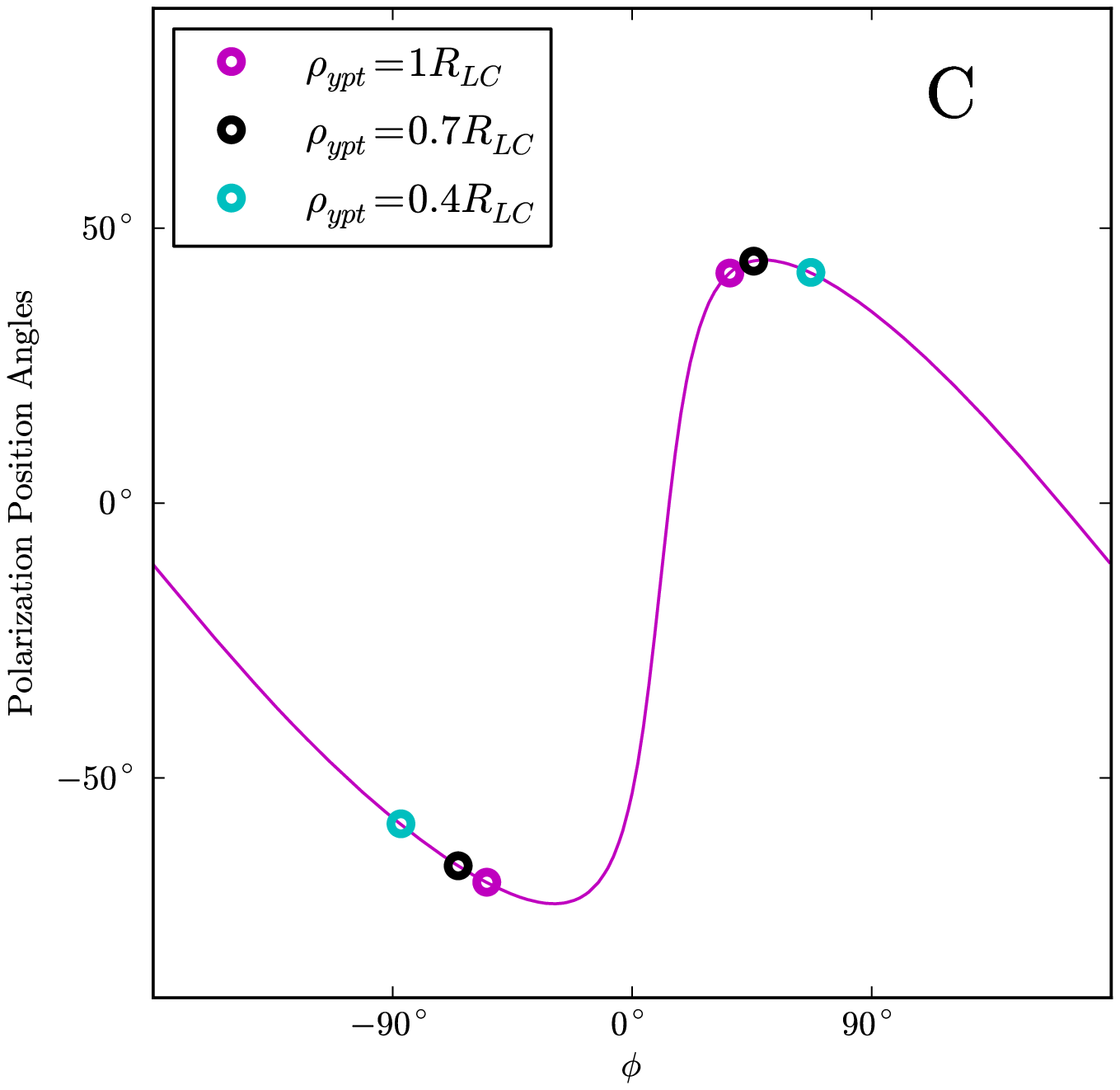}
\includegraphics{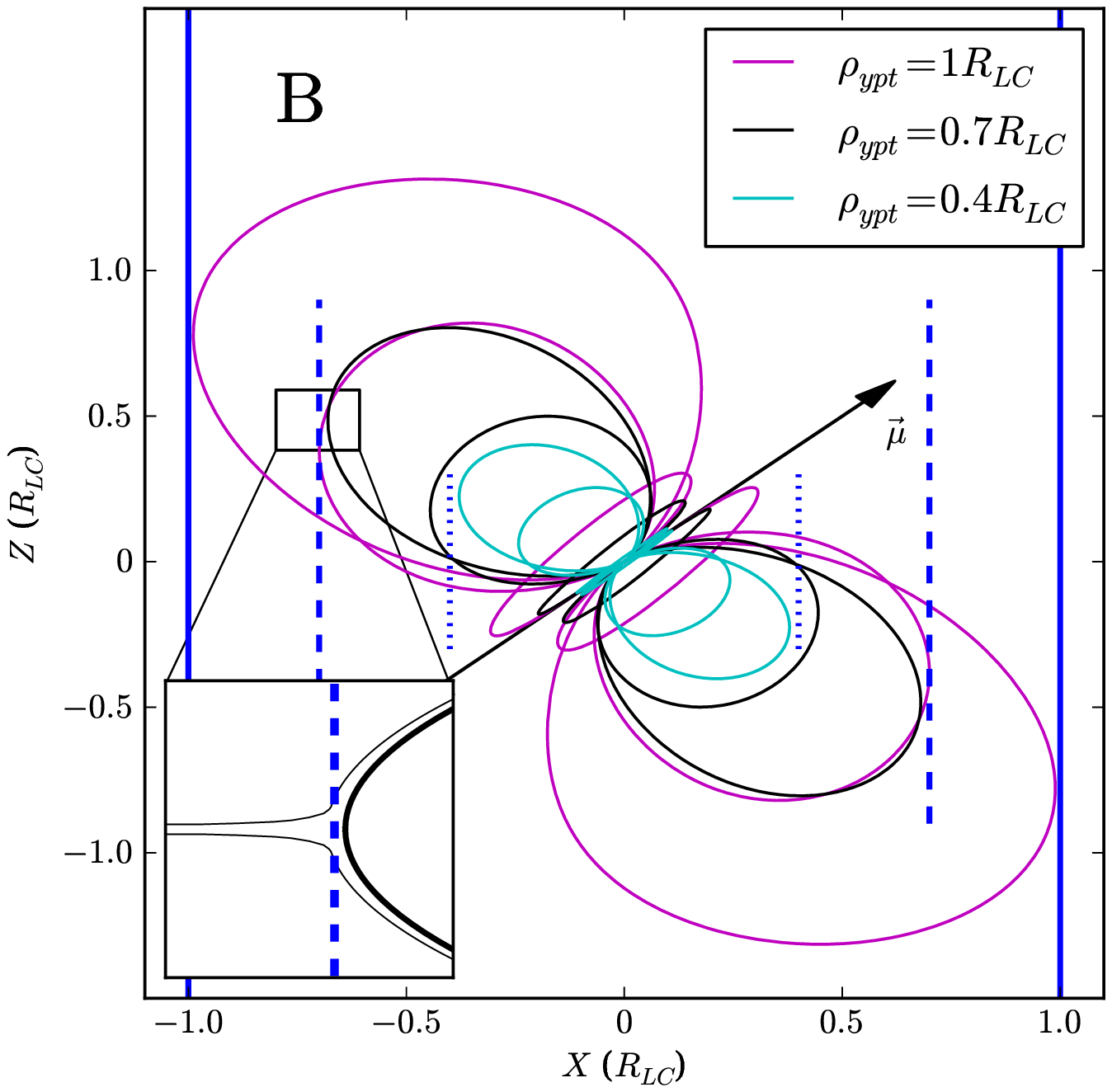}
\begin{center}
\caption{\label{fig:mfl}
Panels (A) and (B) show pulsar magnetic field lines at various viewing angles for a pulsar with $\alpha=45^\circ$.
The magenta lines are the last closed field
lines of a vacuum dipole and the solid blue lines represents the light
cylinder ($1 R_{\rm{LC}}$).  Formally, for the vacuum dipole model, the
y-point radius ($\rho_{\rm{ypt}}$), the cylindrical radius from the spin
axis at which the closed and open field lines
are adjacent, is $\rho_{\rm{ypt}}=1 R_{\rm{LC}}$.  In reality, due to finite mass and current
effects, $\rho_{\rm{ypt}}<1 R_{\rm{LC}}$.  Also plotted is the last closed field lines for
$\rho_{\rm{ypt}}=0.7 R_{\rm{LC}}$ and $\rho_{\rm{ypt}}=0.4 R_{\rm{LC}}$.  The inset plot of panel (B) shows
a close-up of the y-point area and illustrates
why this point is called the y-point.  Panel (C) shows a typical model polarization sweep with $\alpha=145^\circ$,
$\zeta=140^\circ$, and $R=0.1R_{\rm{LC}}$.  Open circles mark the expected emission phase for the model.  
Decreasing $\rho_{\rm{ypt}}$ increases the phase of emission.   A phase of zero is the point
of closest encounter to the magnetic axis in the model.
}
\end{center}
\vskip -0.2truecm
\end{figure*}

The formula is approximate and breaks down as altitude increases.  The
breakdown occurs between $\sim 0.05R_{\rm{LC}}$ and $\sim 0.12R_{\rm{LC}}$
at best, below or near altitudes expected for energetic young and
millisecond pulsars.  One can apply correction
formulae to boost the break-down altitude to $\sim 0.3R_{\rm{LC}}$,
but even these formulae are sensitive to $\alpha$ and $\zeta$
and depend on an assumed radio intensity model (\citealp{craig2012altitude}).
In essence, neither the RVM nor the BCW captures the morphological changes
in the radio polarization position angle sweep at
high altitudes which are needed to model high-energy, $\gamma$-ray emitting pulsars.

Further, the RVM produces smooth S-shaped position angle sweeps
versus phase that resemble data from older, low-energy
pulsars.  Polarization position angle data from millisecond pulsars,
on the other hand, are riddled with jumps, cusps, and sharp turns
(i.e., \citealp{yan2011polarization}; \citealp{everett2001emission})
which are absent in the RVM (the strongest
evidence that the RVM lacks essential physical features
needed to understand emission from these pulsars).  By combining
four physically motivated, data-driven ingredients (numerically calculated
finite altitude, multiple altitudes, orthogonal mode jumps, and interstellar
scattering), we hope to explain some of the features seen in the radio data
of \textit{Fermi} pulsars for which the RVM alone fails.

In the model presented in this paper, and in contrast to the RVM and the BCW, 
finite altitude polarization is calculated using
numerical computation, which avoids the approximations needed in the BCW. 
The retarded dipole presented in \cite{kaburaki1980determination} and used in \cite{Watters:2010jb}
is used in this modeling.  Particles follow magnetic field lines
to a given altitude of emission (as measured radially from the center of the 
neutron star) and, similar to the RVM, radiate tangent to the
field line.  Co-rotation and time-of-flight effects are then applied when emission from the lab
frame is projected onto the plane of the sky 
(that is, $\psi=\psi_v+(|\boldsymbol{\Omega}|/c) \boldsymbol{r}\cdot\hat{k}$
where $\psi$ is the pulsar phase, $\psi_v$ is the co-rotational velocity in the $\hat{\psi}$ direction, $\boldsymbol{r}$ is the 
origin of the emitted photon, and $\hat{k}$ is the direction of the photon motion in the co-rotation frame).  The numerical model is particularly valid
at lower altitudes of emission and we will often favor fits to the model with
low altitude results.  High altitude emission requires a force-free model.  We ignore magnetospheric charge and current present in force-free
models, implicitly assuming that such effects occur at higher altitudes than considered here.

The numerical model does not include the superposition of multiple emission heights which
would result in caustics; the altitudes are defined by a single radial distance from the center
of the neutron star.
The numerical model does not include cross-drift of particles
or higher-order multipoles.

\section{adding physical ingredients}
\label{sec:adding}
\subsection{Multiple Altitudes}
\label{sec:MASE}
In order to include multiple altitudes, 
we invoke the patchy cone model \citep{lyne1988shape,karastergiou2007empirical},
which holds that different components of the pulsar intensity profile come from
different areas of the magnetosphere, and hence, different altitudes.
The polarized intensity is modeled by a combination of Gaussian profiles.
The Gaussian profiles are modeled after those used in \cite{karastergiou2009complex}. 
Each Gaussian component or set of Gaussian components is assigned an altitude
($R_1$, $R_2$, etc.).  We used as few altitudes as will result in a reasonable
fit and multiple components often have a single altitude.  
The Stokes parameters are calculated from the model polarization position angles
and these Gaussian 
components weigh the Stokes parameters $Q$ and $U$ from different
altitudes to calculate a single polarization position angle per phase bin:
\begin{equation}\label{eq:stokes}
\begin{array}{l}
Q_{\rm{tot}}(\phi)=\sum\limits_{n} g_{n}(\phi) \cos{(2\psi_{n}(\phi))},\\\\
U_{\rm{tot}}(\phi)=\sum\limits_{n} g_{n}(\phi) \sin{(2\psi_{n}(\phi))},\\\\
\psi(\phi)=\frac{1}{2}\arctan\left(\frac{U_{\rm{tot}}(\phi)}{Q_{\rm{tot}}(\phi)}\right)
\end{array}
\end{equation} \citep{karastergiou2009complex}.
Here $g_{n}(\phi)$ is a Gaussian component and $\psi_{n}(\phi)$ is the model polarization
associated with that component.

The allowable altitude range in the model is $R=R_{\rm{NS}}$ 
(the neutron star radius) to $R=.9R_{\rm{LC}}$.
Admittedly, we do not attempt to quantify how high of an emission height is 
too far from the neutron star surface to apply a vacuum model for predictions of polarization position angles. 
In all likelihood, there will be a smooth deviation of
the polarization predicted
by the vacuum model 
from the actual polarization
with increasing model altitude and with strong dependence on
$\alpha$, $\zeta$, and the exact location of the emission origin within the magnetosphere. 
Recent and future studies combining vacuum and force-free models (\citealp{kalapotharakos2012gamma})
applied to polarization may hold the key to quantifying this breakdown.

The cone--core model is similar to the patchy cone model but more restrictive.
Physically, a cone of emission beams from the pulsar cap,
flaring out at higher altitudes \citep{radhakrishnan1969magnetic}.
The central part of the intensity pulse profile originates from emission 
low in the magnetosphere near the
neutron star surface and the wings of the intensity pulse profile
originates from emission high in the magnetosphere.
We do not force a cone--core model when fitting altitude but radio modeling 
of J0023$+$0923 and J1024$-$0719 favor high altitude--low altitude--high altitude emission (versus phase)
based on polarization fitting as discussed in Sections~\ref{sec:J0023} and
~\ref{sec:J1024}.

The number of model altitudes used in each fit was motivated
by the polarization data.  For J0023$+$0923 and J1024$-$0719,
we applied two altitude fits since our conjecture is that the 
``jump'' seen in the polarization sweep is from a change
in emission altitude
(Sections~\ref{sec:J0023} and ~\ref{sec:J1024}).
For J1057$-$5226 and J1744$-$1134 
(Sections~\ref{sec:J1057} and~\ref{sec:J1744}), we applied both one and
two altitude fitting schemes.  Although a single altitude
fit would result in a simpler model, it is not unreasonable
to assume that emission from opposite poles or at drastically 
different pulsar phases originates from different heights.
Only one altitude was used in the fitting of the polarization position
angle data of J1420$-$6048 because of the single smooth sweep
in the data (Section~\ref{sec:J1420}).  Polarization data from J2124$-$3358 
(Section~\ref{sec:J2124}) was
fit with more altitudes than reported here but 
such fits did not drastically change the $\chi_{\rm{min}}^2$, the $\chi^2$
map, nor the fit altitudes.  For the sake of simplicity, we 
only report the three altitude fit results.

\subsection{Interstellar Scattering}
Interstellar scattering causes a delay of signal as it travels through
the medium of space.  The result is a delay of the peak, an exponential
tail on the intensity profile, and a flattening of the position angle sweep as polarization information from
earlier phases ``leaks" into polarization in later phases (e.g., \citealp{li2003effect}).
Scattering can be characterized by a scattering time constant and
a scattering kernel:
\begin{equation}\label{eq:gts}
g_{ts}(t-t')=\left\{
\begin{array}{lr}
0,                       & t-t' < 0, \\
e^{-(t-t') / \tau_s},  & t-t' > 0
\end{array}
\right.
\end{equation}
\citep{cronyn1970analysis}.
Other response functions also exist, but this scattering kernel
(a thin scattering screen halfway between the source and 
the observer) is incorporated in the model of this paper.
Scattering time constants ($\tau_s$) as calculated using the \cite{cordes2002ne2001} model are
used in the computations but are negligible for all pulsars
except for J1420$-$6048 in which scattering time was
a free parameter (see Section~\ref{sec:J1420} for details).

Adding scattering to position angle polarization is done
by convolving the scattering kernel with $Q_{\rm{tot}}(\phi)$ and 
$U_{\rm{tot}}(\phi)$:
\begin{equation}\label{eq:con}
\begin{array}{l}
Q_{\rm{tot}}^{\rm{scat}}(\phi)=\int Q_{\rm{tot}}(\phi(t')) g(t-t') dt',
\\ \\
U_{\rm{tot}}^{\rm{scat}}(\phi)=\int U_{\rm{tot}}(\phi(t')) g(t-t') dt'.
\end{array}
\end{equation}
The resulting $Q_{\rm{tot}}^{\rm{scat}}(\phi)$ 
and $U_{\rm{tot}}^{\rm{scat}}(\phi)$ are plugged
into Equation (\ref{eq:stokes}) (bottom line) to obtain polarization.  Further, 
model linear intensity with interstellar scattering is calculated using 
$\sqrt{Q_{\rm{tot}}^{\rm{scat}}(\phi)^2+U_{\rm{tot}}^{\rm{scat}}(\phi)^2}$.

\subsection{Orthogonal Mode Jumps in the Context of Multiple Altitudes and Interstellar Scattering}
\label{sec:MAISOM}
The model also includes orthogonal mode jumps in the polarization position
angle sweep \citep{backer1976orthogonal}. 
To create orthogonal jumps, $90^\circ$ is added to the model 
polarization position angles ($\psi+90^\circ$); Q and U model values can be
calculated and used in Equations (\ref{eq:stokes}) and (\ref{eq:con}) the
same as unjumped polarization.
We do not attempt to understand the origin of
these jumps since our model does not contain the physics needed to do so
but rather we use the jumps on the basis of empirical observation (e.g.,
\citealt{stinebring1984pulsar}; \citealt{gould1998multifrequency}; \citealt{karastergiou2005polarization}).
Note that components of polarized intensity at the same altitude but in different
modes will cancel exactly where their individual absolute intensities are equal (Equation (\ref{eq:stokes})).
Polarized intensity going to zero at a given phase in the intensity profile
indicates an orthogonal mode jump.  Further, if the orthogonal mode 
jump occurs between components of different altitudes, the polarized intensity 
will not cancel exactly, although in most cases it will be near zero.

Additionally, without multiple altitudes, only $90^\circ$ jumps are allowed
and the direction of the bridging polarization position angle sweep between the jump is in the opposite direction from the sweep
without the jump.
Mathematically, this is due to the forward scattering nature
of the scattering kernel.  Equation (\ref{eq:con}) is nonzero
only for $t-t'>0$ and convolution with such a function
will mathematically cause polarization from earlier in phase to mix
with polarization at any given phase.
Similarly, this is why the second model Gaussian component
in phase ($C_2$) is higher in intensity than the first ($C_1$)
in Figure~\ref{fig:PlotJexampleintPA}. 
 
Figure~\ref{fig:PlotJexampleintPA} illustrates some properties
of scattering and mode jumps that will be of particular interest in
the analysis of fitting J0023$+$0923 and J2124$-$3358 polarization data.  An orthogonal mode jump
in the position angle sweep between components of the same altitude
will result in a bridging sweep with the opposite curvature compared
with the unjumped sweep.  In the example figure, the magenta solid line
has an upward curvature at the jump phase.  The resulting curve with the
addition of an orthogonal mode jump (black dots) curves downward at the jump
phase.  This opposite curvature should always occur
due to forward scattering if the individual component altitudes
are exactly the same.  If the component altitudes are not
the same, the bridging curvature between the orthogonal polarization position
angles could be the same as or opposite the curvature of the original sweep
direction depending on the polarization position angles (and Stokes parameters) being
combined.  This will be an important argument for orthogonal
mode jumps between different altitude components in Sections~\ref{sec:J0023}
and~\ref{sec:J2124}.

In Figure~\ref{fig:PlotJexampleintPA} (along with all subsequent graphs with an $x$-axis of 
pulsar phase), the phase of zero is the point of closest encounter to the magnetic axis in 
the model.

\section{Fitting Methodology}
\label{sec:fit}

Simple Gaussian curves were used to model the pulsar intensity. A fixed set
of Gaussian phases, widths, and amplitudes drawn by eye that mimic the 
linear intensity amplitude were used.  
Formally fitting the pulsar intensity
would require simultaneously fitting the polarization position angle parameters and the 
Gaussian parameters.  Such a fit would be computationally intensive and have minimal corrections
to the model fits of the polarization. 

For the fitting of the polarization position angles, we fit the horizontal and vertical offsets 
($\Delta\phi$ and $\Delta\psi$) and altitudes ($R_{1}$, $R_{2}$, etc.) with $\alpha$ and $\zeta$ 
fixed in $1^\circ$ increments.  We used a simulated annealing scheme to find the global minimum for
a given $\alpha$ and $\zeta$ ~\citep{flannery1992numerical}.  
We then randomly sample the surrounding parameter space within $3\sigma$ of the
lowest $\chi^2$ for every fixed $\alpha$--$\zeta$ pair to calculate fit error bars.

Phase cuts were applied to the polarization data points where total normalized intensity dropped below 
$10\%$ for a given pulse. Error bar cuts were also applied to data points where error bars exceeded $\pm20^\circ$.
Error bar cuts were chosen such that we had good confidence that data points 
are within half of the $180^\circ$ range that they can occupy 
(because of the possibility of orthogonal mode jumps).  
Phase cuts were chosen such that only data points with 
a reasonable signal-to-noise ratio were considered.

\subsection{y-point Considerations}
\label{sec:ypt}
The light cylinder is defined as $R_{\rm{LC}}=cP/2\pi$ where
$c$ is the speed of light and $P$ is the period of the pulsar.
This cylindrical distance (measured from the neutron
star rotation axes) is where particles in co-rotation with the
neutron star would be traveling at the speed of light.
At this point (or more physically, before this point), the 
field lines ``break open.''  These open field lines are where
in the magnetosphere particles accelerate and the pulsar radiates. 
The point at which an open field line
is adjacent to a closed field line is
the y-point since the adjacent open and closed field lines
form a Y in the field (see Figure~\ref{fig:mfl}(B) inset
for illustration of the y-point).  Figure~\ref{fig:mfl}
(panels (A) and (B)) plots the field lines of a model pulsar with $\alpha=45^\circ$ 
($\boldsymbol{\mu}$ is the magnetic axis and the spin axis is vertical). 
Panel (A) shows a top view of the light cylinder (solid blue lines)
while panel (B) shows a side view of the light cylinder.  The magenta
field lines represent the last closed field lines of a vacuum dipole
model with the y-point occurring just beyond the light cylinder. 

Studies with
force-free simulations \citep{spitkovsky2006time} valid at heights near
the light cylinder indicate the y-point typically occurs
further in than the light cylinder due to particle mass and 
charge current.  The location
of the y-point controls the size of the cap of emission from 
which the open field lines and emission originate.  Smaller $\rho_{\rm{ypt}}$,
the cylindrical distance of the y-point from the neutron star, results in a larger
cap, a wider range of viewing angles over which emission can be seen, and a wider phase over which
emission is produced.  Often, the emission phase in data
is too large to be accommodated by the open
zone of the formal vacuum dipole even with finite altitude;
this is evidence that the field lines break open further in from
the light cylinder. The cyan
and black field lines in Figure~\ref{fig:mfl} illustrate the location and form
of the last closed field lines with the y-point distance equal to
$\rho_{\rm{ypt}}=0.4R_{\rm{LC}}$ or $\rho_{\rm{ypt}}=0.7R_{\rm{LC}}$ respectively.

Panel (C) of Figure~\ref{fig:mfl} shows the effect of a shifted y-point on the range of emission allowed from open field lines.
Panel (C) is a plot of a typical model polarization position angle sweep with $\alpha=145^\circ$,
$\zeta=140$, and $R=0.1R_{\rm{LC}}$.  The open circles mark the region in phase where emission is allowed
for various $\rho_{\rm{ypt}}$.  As $\rho_{\rm{ypt}}$ decreases, the allowed range of emission increases.

In the modeling for this paper, emission 
is treated as coming from all field lines not just those defined as open by the formal
cap using the light cylinder distance.  Polarization data are fit 
without constraints from the emission phase.
We then report the $\rho_{\rm{ypt}}$
needed for the entire phase of emission seen in the intensity data
to be covered by model data in the same phase.

\section{Application (and illustration) with Individual Pulsars}
\label{sec:app}
\subsection{J0023$+$0923: Nonorthogonal Jumps with Multiple Altitudes}

\label{sec:J0023}


\begin{table*}[t!!]
\small
\caption{Fit Parameters for J0023$+$0923}
\tabcolsep=0.19cm
\begin{center}
\begin{tabular}{lccccccc}
\hline
\hline 
& DOF & (Unreduced) $\chi^2_{\rm{min}}$ & $\alpha$ ($^\circ$) & $\zeta$ ($^\circ$)& $R_1$ ($R_{LC}$)& $R_2$ ($R_{LC}$)& $\Delta R$ ($R_{LC}$)\\
[.3em]\hline 
RVM &131-4 &$   1053 $ & $8 ^{+25(+56)}_{-6(-7)} $ & $13 ^{+38(+71)}_{-10(-11)} $ & \nodata& \nodata&\nodata    \\ \hline

Region A 2 Alt & 131-6& $313$ & $57 ^{+15(+53)}_{-14(-30)}$ & $105 ^{+14(+36)}_{-16(-47)} $ 
& $ 0.67 ^{+0.23(+0.23)}_{-0.24(-0.43)} $ & $ 0.28 ^{+0.24(+0.33)}_{-0.14(-0.20)} $ 
& $ 0.39 ^{+0.23(+0.35)}_{-0.11(-0.23)} $   \\ \hline

Region B 2 Alt & 131-6&$   324 $ & $   59 ^{+8(+28)}_{-10(-58)} $ & $   49 ^{+2(+7)}_{-5(-48)} $ 
& $ 0.53 ^{+0.07(0.17)}_{-0.07(-0.22)} $ & $ 0.90 ^{+0.00(+0.00)}_{-0.06(-0.27)} $ 
& $-0.37 ^{+0.08(+0.17)}_{-0.06(-0.22)} $  \\ \hline
\end{tabular}
\tablecomments{Errors reported without (with) parentheses are for $1\sigma$ ($3\sigma$) from $\chi^2_{\rm{min}}$. }
\label{tb:fitJ0023}
\end{center}
\end{table*}

\begin{figure}[t!!]
\vskip .39\textheight
\includegraphics{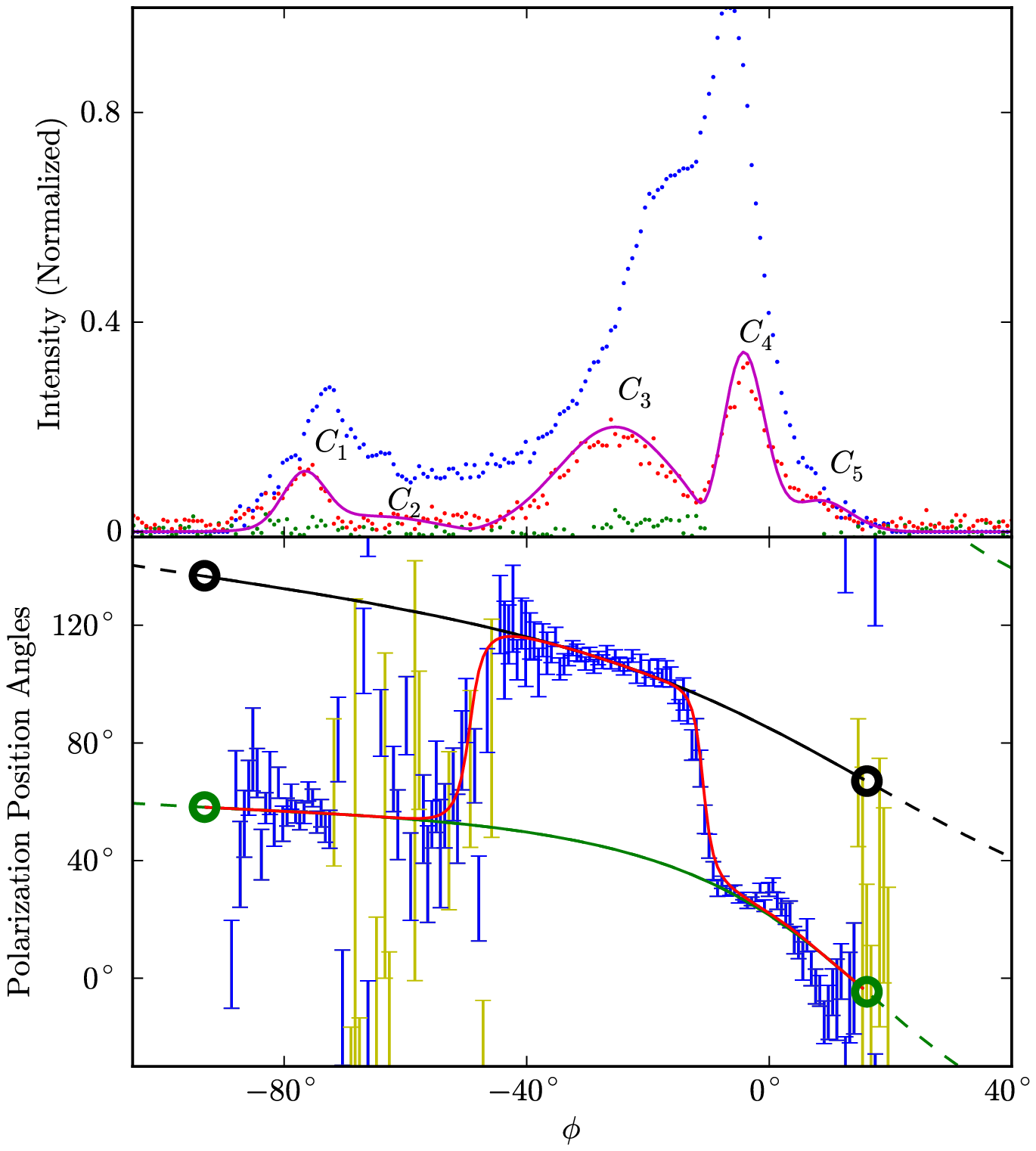}
\begin{center}
\caption{\label{fig:PlotJ0023intPA}
In the upper panel, blue points are total radio intensity data at 1.646 GHz, red points are linear polarization intensity data
and green points are circular polarization intensity data for J0023$+$0923.  The solid magenta line in the upper panel is
the model linear intensity used in fitting.  In the bottom panel, blue error bars are polarization position angles
used in the fit and yellow error bars are polarization position angles excluded by error bar cuts (but not 
excluded by phase cuts).  The model polarization comes from
a fit with (unreduced) $\chi^2=332$ and parameters $\alpha=59^{\circ}$ and $\zeta=119^{\circ}$.  The green solid line is the polarization for
a model with $R_{1}=0.77R_{\rm{LC}}$ and the black solid line is the polarization for 
a model with $R_{2}=0.38R_{\rm{LC}}$.  The red solid line is the model
polarization of the two altitudes weighted by the model intensity. Empty circles mark the limiting phase of emission from
open field lines with $\rho_{\rm{ypt}}=1R_{\rm{LC}}$. A phase of zero is the point
of closest encounter to the magnetic axis in the model.
}
\end{center}
\vskip -0.2truecm
\end{figure}

\begin{figure}[t!!]
\vskip .44\textheight
\includegraphics{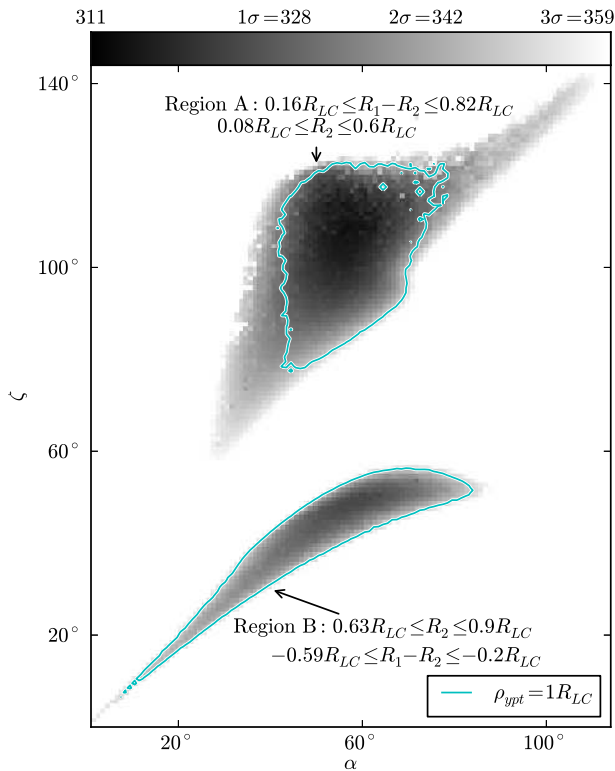}
\begin{center}
\caption{\label{fig:J0023Map}
Map of (unreduced) $\chi^{2}$ for J0023$+$0923 in the $\alpha$--$\zeta$ plane.  Cyan contours mark $3\sigma$ from $\chi^2_{\rm{min}}$ 
for fits with $\rho_{\rm{ypt}}=1R_{\rm{LC}}$.  This contour contains the lowest values of $\chi^2$.  The two
regions of statistically acceptable fits have drastically different fit parameters.
}
\end{center}
\vskip -0.2truecm
\end{figure}

J0023$+$0923 is a \textit{Fermi} millisecond pulsar with $P=3.05$ ms.  Figure~\ref{fig:PlotJ0023intPA}
shows the radio pulse profile and the polarization position angles at 1.646 GHz.
The polarization sweep cannot be explained well using the RVM because of
sharp curvature between intensity components $C_2$ and $C_3$ 
and between components $C_3$ and $C_4$.  
The RVM with an orthogonal mode jump between these components
produces more reasonable fits as reported in Table~\ref{tb:fitJ0023}.
This fit is unsatisfying because the jump is closer to $\sim 60^\circ$ rather
than $90^\circ$.  Fitting with an orthogonal mode jump plus two 
altitudes (one altitude, $R_{1}$, assigned to $C_1$, $C_2$, $C_4$, and $C_5$
and a second altitude, $R_{2}$, assigned to $C_3$) gives a fit 
with significantly smaller $\chi^2_{\rm{min}}$ 
(unreduced $\chi^2_{\rm{min}}=313$ versus unreduced $\chi^2_{\rm{min}}=1053$). 
By including two physically motivated
parameters (the two altitudes), the $\chi^2_{\rm{min}}$ is decreased by a factor of three.
The $F$-test between the RVM and the two-altitude model gives $F=149.12$, $\rm{DOF}_1=2$, and $\rm{DOF}_2=125$. The probability of
exceeding this $F$ is $\rm{Prob}\sim 0$.  This pulsar is an excellent
example of how modeling with multiple altitudes can greatly improve $\chi^2_{\rm{min}}$ compared to the RVM.
It is an example of how multiple altitudes can easily explain a non-$90^\circ$ orthogonal mode jumps since
components of emission with different altitudes allow
for these non-$90^\circ$ mode jumps.
Figure~\ref{fig:PlotJ0023intPA} shows the polarization position angle and intensity data
overlaid with the best fit two altitudes plus the orthogonal mode jump model.

Further confirmation that this mode jump is between components from different
altitudes in the magnetosphere comes from the curvature direction of the polarization position 
angle sweep between adjacent modes.
The polarization sweep direction between components $C_3$ and $C_4$ is in the 
same direction as both the individual unweighted model curves (solid black and red lines).
Such a direction of curvature is impossible for a mode jump between equal altitudes as
discussed in Section~\ref{sec:MAISOM}.  Also, such a direction of curvature is impossible
in the RVM model which partially accounts for the poor fit.

Satisfactory fits exist for both $\rho_{\rm{ypt}}= 1 R_{\rm{LC}}$ and $\rho_{\rm{ypt}}<R_{\rm{LC}}$.
Figure~\ref{fig:J0023Map} is the (unreduced) $\chi^{2}$ map in the $\alpha$--$\zeta$ plane and
the thin cyan contour represents the allowable area up to $3\sigma$ from $\chi^{2}_{\rm{min}}$ in
which the emission comes only from the formal open field line region of the magnetic pole 
as defined by $\rho_{\rm{ypt}}=1R_{\rm{LC}}$.
The minimum $\chi^2$ region is well within the $\rho_{\rm{ypt}}=1R_{\rm{LC}}$ region as seen from Figure~\ref{fig:J0023Map}.
Also Figure~\ref{fig:PlotJ0023intPA} shows a polarization position angle model that 
emits over the entire phase of emission seen in the data where the circles on the plot mark the phase defining
the formal open zone for the two altitudes of emission used in the model.

In the $\alpha$--$\zeta$ plane, two islands of acceptable regions of $\chi^2<3\sigma$ arise as seen in Figure~\ref{fig:J0023Map}.  
Parameters and errors for each of these sections are reported separately in Table~\ref{tb:fitJ0023}.
The distinguishing parameter
between these two regions is $R_{2}$.  For the region between $\zeta=1^{\circ}$ and $\zeta=56^{\circ}$,
$R_{2}=0.63$--$0.90R_{\rm{LC}}$.  For the region between $\zeta=58^{\circ}$ and $\zeta=141^{\circ}$,
$R_{2}=R_{\rm{ns}}=0.08$--$0.61R_{\rm{LC}}$.  
The region between $\zeta=58^{\circ}$ and $\zeta=141^{\circ}$
is more plausible for our model because it favors lower altitudes.  
Additionally, $\Delta R=R_{1}-R_{2}$ for
this region is positive and favors the cone--core model discussed previously (see Table~\ref{tb:fitJ0023}
for values).

\subsection{J1024$-$0719: Kinks with Multiple Altitudes}
\label{sec:J1024}

\begin{table*}[t!!]
\footnotesize
\caption{Fit Parameters for J1024$-$0719}
\tabcolsep=0.05cm
\begin{center}
\begin{tabular}{lccccccccc}
\hline
\hline
 &&(Unreduced)&&&&&&&\\
 &DOF&$\chi^{2}_{\rm{min}}$& $\alpha$ ($^\circ$) & $\zeta$ ($^\circ$) & $R_1$ ($R_{\rm{LC}}$)& $R_2$ ($R_{\rm{LC}}$)& $\Delta R$ ($R_{\rm{LC}}$)& $\Delta \rho$ ($R_{\rm{LC}}$)& $\rho_{\rm{ypt}}$ ($R_{\rm{LC}}$)\\
[.3em] \hline  
RVM & 397-4&$  8963 $ & $98 ^{+1(+3)}_{-3(-4)} $ & $87 ^{+1(+2)}_{-2(-2)} $&\nodata &\nodata &\nodata &\nodata &\nodata \\ \hline
Region A 2 Alt &397-6&$  3409 $ & $  112 ^{  + 1(+2)}_{   -1(-3)} $ & $   63 ^{+2(+5)}_{-3(-6)} $ 
& $ 0.90 ^{+0.00(+0.00)}_{-0.02(-0.05)} $ & $ 0.82 ^{+0.00(+0.01)}_{-0.02(-0.05)} $ 
& $ 0.08 ^{+0.01(+0.01)}_{-0.00(-0.01)} $ & $ 0.11 ^{+0.00(+0.01)}_{-0.01(-0.01)} $ & $ 1.00 ^{+0.00(+0.00)}_{-0.02(-0.04)} $ \\ \hline
Region B 2 Alt &397-6&$  3447 $ & $  113 ^{+1(+4)}_{-1(-4)} $ & $  108 ^{+1(+4)}_{-1(-4)} $
& $ 0.22 ^{+0.01(+0.04)}_{-0.03(-0.05)} $ & $ 0.05 ^{+0.00(+0.01)}_{-0.00(-0.00)} $ 
& $ 0.17 ^{+0.01(+0.04)}_{-0.03(-0.05)} $ & $ 0.04 ^{+0.01(+0.08)}_{-0.01(-0.03)} $ & $ 0.24 ^{+0.00(+0.07)}_{-0.02(-0.04)} $ \\ \hline
\end{tabular} 
\tablecomments{Errors reported without (with) parentheses are for $1\sigma$ ($3\sigma$) from $\chi^2_{\rm{min}}$. }
\label{tb:fitJ1024} 
\end{center} 
\end{table*}

\begin{figure}[t!!]
\vskip .39\textheight
\includegraphics{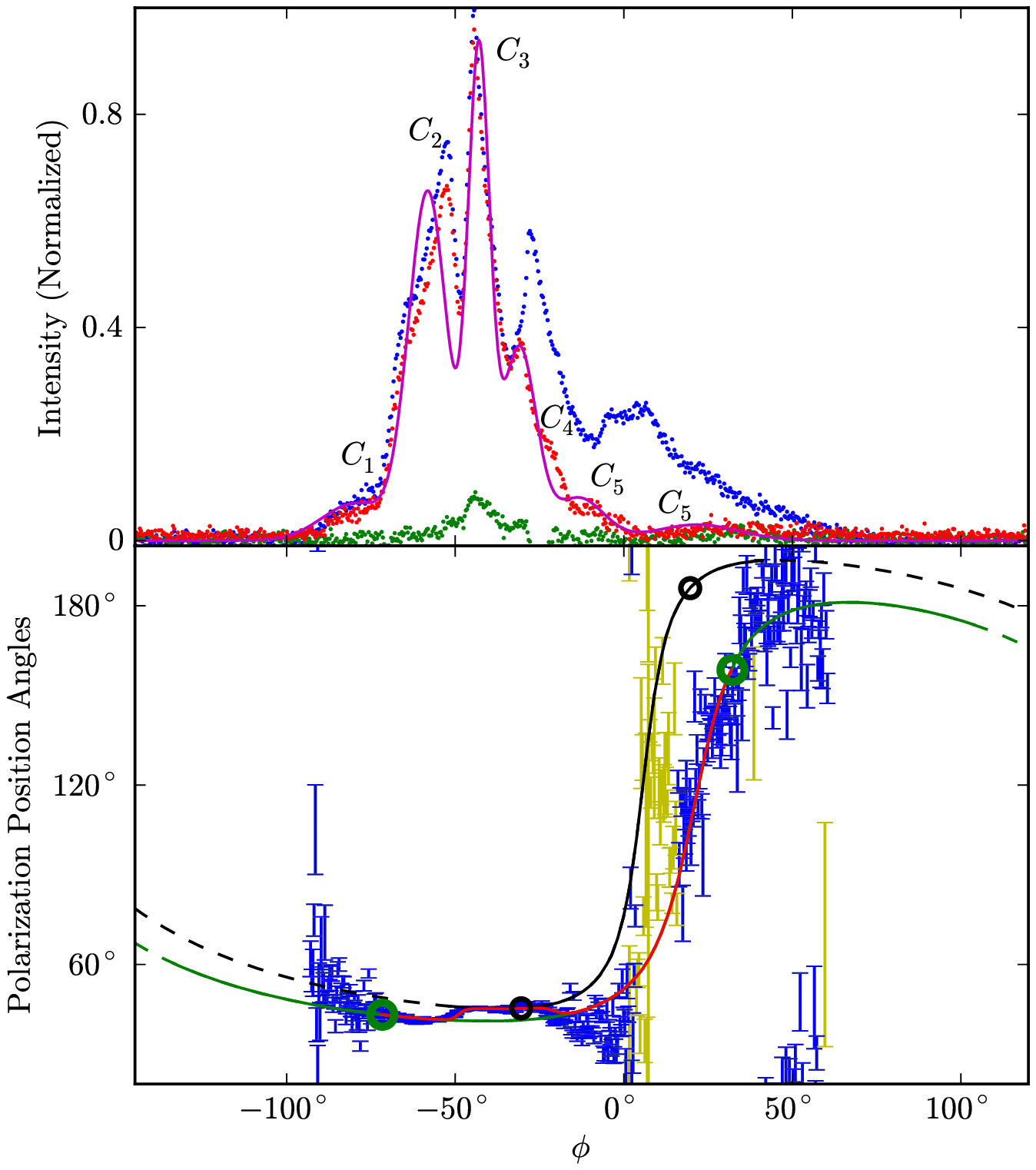}
\begin{center}
\caption{\label{fig:PlotJ1024intPA}
In the upper panel, blue points are total radio intensity data for 1.369 GHz, red points are linear polarization intensity data,
and green points are circular polarization intensity data for J1024$-$0719.  The solid magenta line in the upper panel is
the model linear intensity used in fitting.  In the bottom panel, blue error bars are polarization position angles
used in the fit and yellow error bars are polarization position angles excluded by error bar cuts (but not 
excluded by phase cuts).
The model polarization comes from
a fit with (unreduced) $\chi^2=3448$ and parameters $\alpha=113^{\circ}$ and $\zeta=108^{\circ}$.  The green solid line is the polarization for
a model with $R_{1}=0.22R_{\rm{LC}}$ and the black solid line is the polarization for 
a model with $R_{2}=0.05R_{\rm{LC}}$.  The red solid line is the model
polarization of the two altitudes weighted by the model intensity. Empty circles mark the limiting phase of emission from
open field lines with $\rho_{\rm{ypt}}=1R_{\rm{LC}}$.  
Solid lines mark the allowed emission phase for an effective open zone with $\rho_{\rm{ypt}}=0.24R_{\rm{LC}}$ which is
required for the model phase to cover the entire emission phase in the data.
A phase of zero is the point
of closest encounter to the magnetic axis in the model.
}
\end{center}
\vskip -0.2truecm
\end{figure}

\begin{figure}[t!!]
\vskip .27\textheight
\includegraphics{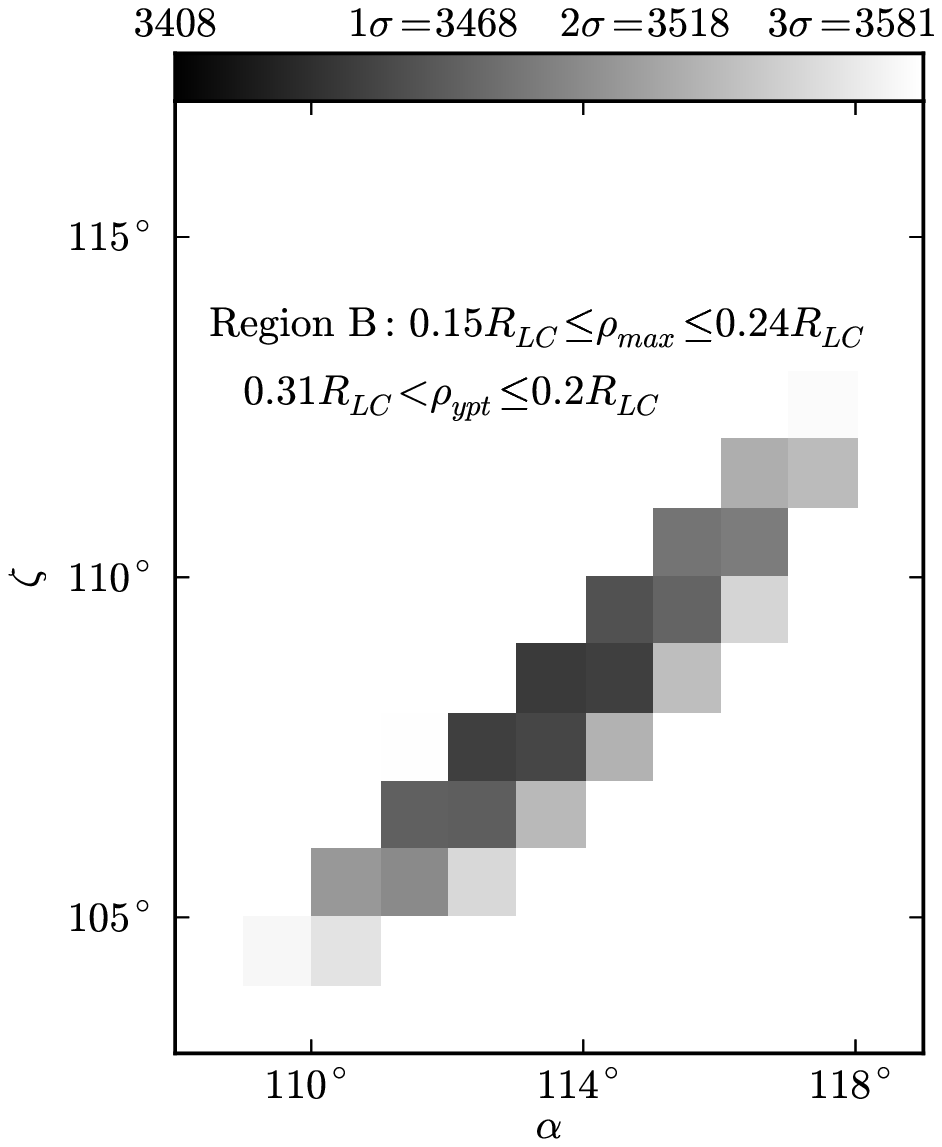}
\includegraphics{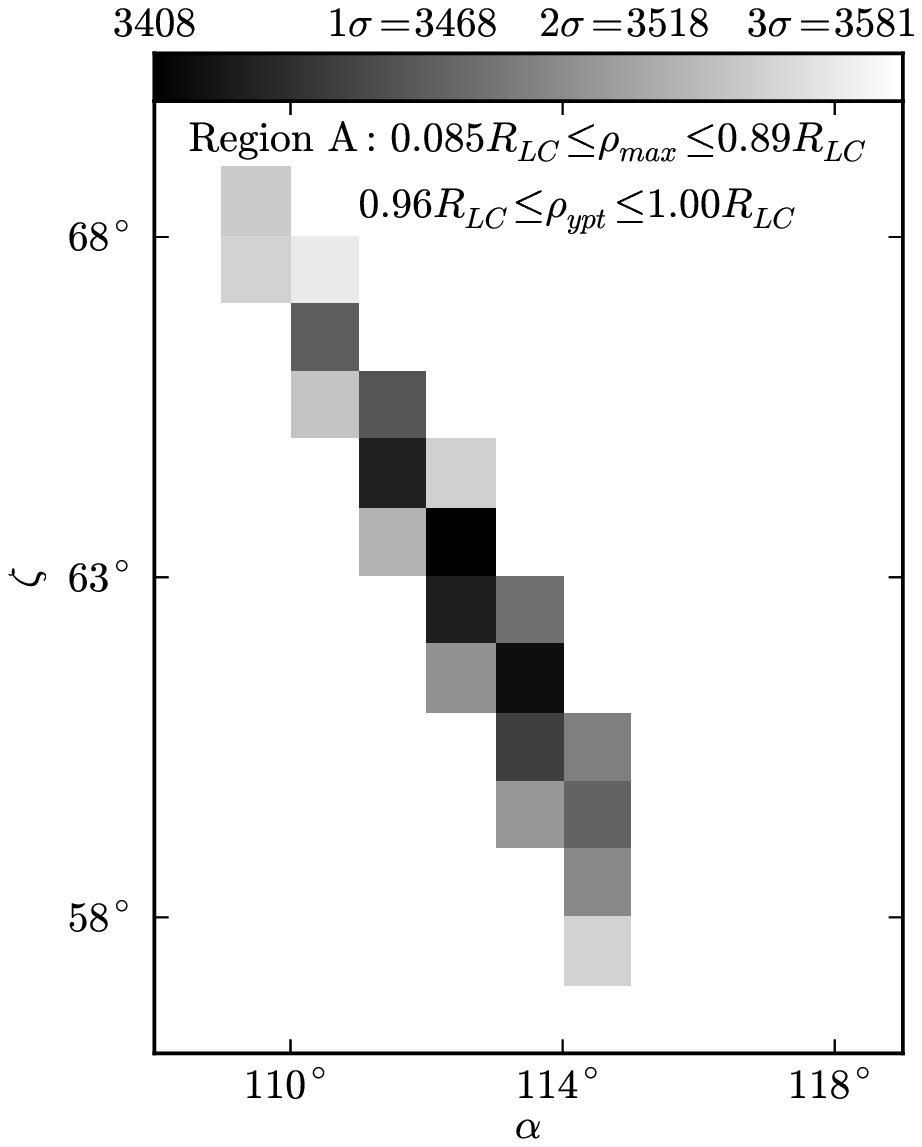}
\begin{center}
\caption{\label{fig:J1024Map}
Map of (unreduced) $\chi^{2}$ for J1024$-$0719 in the $\alpha$--$\zeta$ plane for Regions A and B.
The regions vary drastically from one another in parameters although they are comparable in $\chi^2$.
Both are very restrictive in their respective parameters.  
}
\end{center}
\vskip -0.2truecm
\end{figure}

J1024$-$0719 is another millisecond \textit{Fermi}-detected pulsar ($P=5.162$ ms). 
This pulsar fits reasonably well to the RVM (1.369 GHz data shown in 
Figure~\ref{fig:PlotJ1024intPA}) although certain features 
in the polarization data are highly statistically significant
and unexplained by the RVM.  Most notably, a kink in the polarization occurs in 
the transition from  intensity component $C_2$ to 
$C_3$ and from $C_4$ to $C_5$ (as labeled on Figure~\ref{fig:PlotJ1024intPA}).  
Similar to J0023$+$0923, this jump can be modeled using
multiple altitudes, modeling the kink as the shift in altitude
between the emission components.  The error bars
on this model are fairly small for the fit parameters (Table~\ref{tb:fitJ1024}) 
since a limited number of parameter combinations
make a sweep with this particular polarization difference with two heights.  Additionally, the data contains
the polarization at the point of closest encounter to the magnetic axis (the fastest change in the sweep) which
also greatly constrains the fitting parameters.
Overall, this pulsar is another excellent example of how using multiple altitudes can explain 
features not found in the RVM.

Similar to J0023$+$0923, two regions arise in the $\alpha$--$\zeta$ plane of the (unreduced) $\chi^2$ map
(see Figure~\ref{fig:J1024Map}) that are statistically acceptable.  For the 
two regions, the altitudes are significantly different and are reported separately in Table~\ref{tb:fitJ1024}.

In Region A, the $\rho_{\rm{ypt}}$ needs to be 
$ 0.22 R_{\rm{LC}} \leq \rho_{\rm{ypt}} \leq 0.34 R_{\rm{LC}}$ in order to account for the full 
phase of the emission and remain within $3\sigma$ of $\chi^2_{\rm{min}}$. 
Figure~\ref{fig:PlotJ1024intPA} displays the model polarization sweep from this region.
The green and black circles represent the limiting phase 
of emission defined by the formal cap with $\rho_{\rm{ypt}}=1R_{\rm{LC}}$.
The solid lines mark the effective open zone emission for $\rho_{\rm{ypt}}=0.24R_{\rm{LC}}$.
Both sets of circles are well within the limiting phase of emission seen in the data.
For most of the models within $3\sigma$ of $\chi^2_{\rm{min}}$, 
the emission phase of the outer altitude ($R_1$ represented 
by the green line in Figure~\ref{fig:PlotJ1024intPA}) 
controls the location of $\rho_{\rm{ypt}}$.
For Region A, $R_{1}=0.14$--$0.26R_{\rm{LC}}$, $R_{2}=0.05$--$0.06R_{\rm{LC}}$. 
The outer emission altitude, $R_{1}$, is typically larger than the inner emission altitude, $R_{2}$, which is
consistent with a cone--core model similar to the two fit altitude parameters for the polarization
position data from J0023$+$0923 .

For Region B, the $\rho_{\rm{ypt}}$ is much higher than that in Region A but likewise so are $R_{1}$ and $R_{2}$.
Approximately $\rho_{\rm{ypt}}=0.96$--$1.0R_{\rm{LC}}$, 
$R_{1}=0.85$--$0.9R_{\rm{LC}}$, $R_{2}=0.77$--$0.83R_{\rm{LC}}$.
Although $\rho_{\rm{ypt}}$ is much larger in Region B than in Region A, the altitude of 
emission is also close to $R_{\rm{LC}}$ resulting in a $\Delta \rho=\rho_{\rm{ypt}}-\rho_{\rm{max}}$
(the
difference between the y-point cylindrical distance needed to open up the field lines
to the appropriate amount to accommodate the emission phase in the data and
the maximum cylindrical distance of the emission within this phase)
similar to that of
Region A.  Additionally, $\Delta \rho$
is relatively
small (see Table~\ref{tb:fitJ1024}).  This is problematic because we expect emission 
this close to the light cylinder to resemble a force-free model and to be dictated
by physics that we do not include in the vacuum model.
To explain the phase of emission seen in the data using this model, one must either push the 
altitude up to extreme heights or accept $\rho_{\rm{ypt}}$ much smaller than $1R_{\rm{LC}}$.

J1024$-$0719 is modeled with two altitudes, one altitude component inside the other
in terms of phase, making it a good candidate for the cone--core model.  The parameter
$\Delta R=R_1-R_2$ (where $R_1$ is the altitude associated with component $C_1$, $C_5$, and $C_6$ and
$R_2$ is the altitude associated with $C_2$ and $C_3$) should be positive in the
case of a cone--core model.  Table~\ref{tb:fitJ1024} shows this value to be positive
for both Regions A and B.  
Further, to $3\sigma$, these values are positive and radio modeling of J1024$-$0719 polarization
is consistent with the cone--core model.

For the RVM (unreduced) $\chi^2_{\rm{min}}=8963$ 
and for the two-altitude model (unreduced) $\chi^2_{\rm{min}}=3408$. By including two physically motivated 
parameters, the $\chi^2_{\rm{min}}$ is decreased by a factor of three.  
The $F$-test between the RVM and the two-altitude model gives $F=318.66$, DOF$_1=2$, and DOF$_2=391$. The probability of
exceeding this $F$ is Prob$\sim 0$.  By these statistical measures, a two-altitude model is clearly
better than the RVM.

\subsection{J1057$-$5226: y-point and Finite Altitude}
\label{sec:J1057}

\begin{table*}[t!!]
\small
\tabcolsep=0.05cm
\caption{Fit Parameters for J1057$-$5226} 
\begin{center}
\begin{tabular}{lccccccccc}
\hline
\hline
 &&(Unreduced)&&&&&&&\\
& DOF&$\chi^2_{\rm{min}}$ & $\alpha$ ($^\circ$) & $\zeta$ ($^\circ$) & $R_1$ ($R_{\rm{LC}}$)& $R_2$ ($R_{\rm{LC}}$)& $\Delta \rho$ ($R_{\rm{LC}}$)& $\rho_{\rm{max}}$ ($R_{\rm{LC}}$)& $\rho_{\rm{ypt}}$ ($R_{\rm{LC}}$)\\ 
[.3em]\hline 

RVM & 166-4&$   325 $ & $   77 ^{     +0(+1)}_{   -1(-1)} $ & $   70 ^{     +0(+0)}_{    -0(-0)} $  &\nodata &\nodata &\nodata &\nodata &\nodata \\ \hline

1 Alt & 166-5&$   282 $ & $   76 ^{+2(+2)}_{0(-1)} $ & $   64 ^{+5(+5)}_{-34(-35)} $  & $ 0.19 ^{+0.43(+0.46)}_{-0.12(-0.13)} $ &\nodata  &\nodata &\nodata &\nodata\\ \hline

2 Alt & 166-6&$   281 $ & $   69 ^{+14(+26)}_{-4(-9)} $ & $   64 ^{+5(+6)}_{-37(-39)} $ 
& $ 0.12 ^{+0.72(+0.72)}_{-0.06(-0.12)} $ & $ 0.31 ^{+0.59(+0.59)}_{-0.29(-0.31)} $ 
& $ 0.15 ^{+0.32(+0.40)}_{-0.15(-0.15)} $ & $ 0.31 ^{+0.36(+0.52)}_{-0.25(-0.28)} $ & $ 0.46 ^{+0.54(+0.54)}_{-0.40(-0.40)} $ \\ \hline

\end{tabular} 
\tablecomments{Errors reported without (with) parentheses are for $1\sigma$ ($3\sigma$) from $\chi^2_{\rm{min}}$. }
\label{tb:fitJ1057}
\end{center} 
\end{table*}

\begin{figure}[b!!]
\vskip .39\textheight
\includegraphics{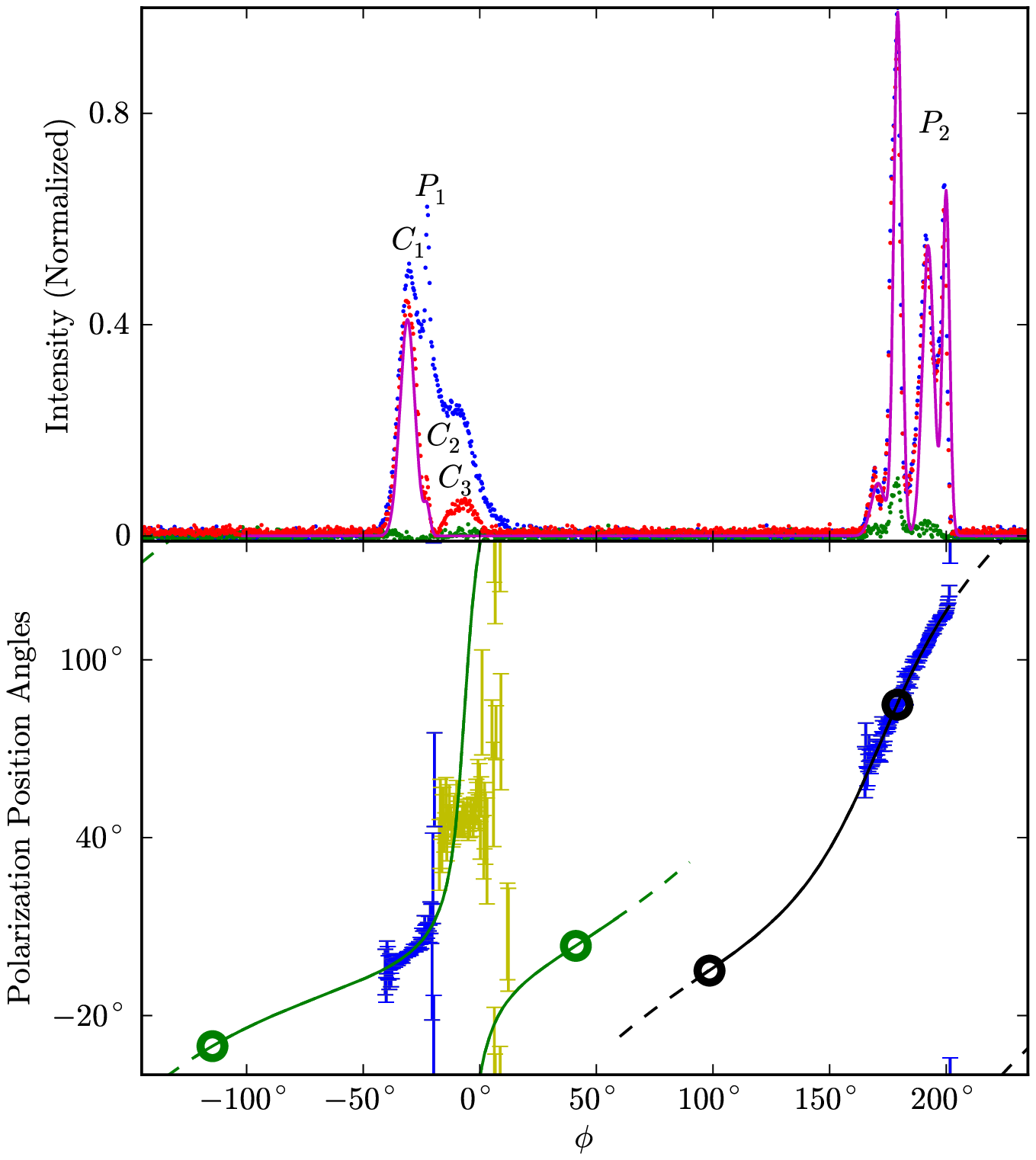}
\begin{center}
\caption{\label{fig:PlotJ1057intPA}
In the upper panel, blue points are total radio intensity data for 1.5 GHz, red points are linear polarization intensity data,
and green points are circular polarization intensity data for J1057$-$5226.  The solid magenta line in the upper panel is
the model linear intensity used in fitting.  In the bottom panel, blue error bars are polarization position angles
used in the fit and yellow error bars are polarization position angles excluded from the 
fit because of phase cuts (see text).
The model polarization comes from
a fit with (unreduced) $\chi^2=289$ and parameters $\alpha=77^{\circ}$ and $\zeta=30^{\circ}$.  The green solid line is the polarization for
a model with $R_{1}=0.58R_{\rm{LC}}$ and the black solid line is the polarization for 
a model with $R_{2}=0.63R_{\rm{LC}}$.  The red solid line is the model
polarization of the two altitudes. Empty circles mark the limiting phase of emission from
open field lines with $\rho_{\rm{ypt}}=1R_{\rm{LC}}$.
Solid lines mark allowed emission phase for an effective open zone with
$\rho_{\rm{ypt}}=0.71R_{\rm{LC}}$ ($\Delta\rho=0.49R_{\rm{LC}}$) which is
required for the model phase to cover the entire emission phase in the data.
A phase of zero is the point
of closest encounter to the magnetic axis in the model.
}
\end{center}
\vskip -0.2truecm
\end{figure}

\begin{figure}[t!!]
\vskip .41 \textheight
\includegraphics{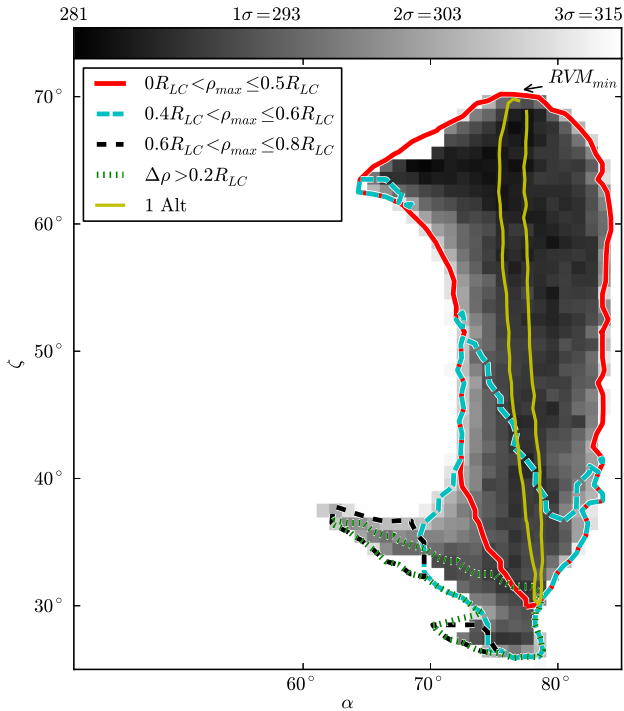}
\begin{center}
\caption{\label{fig:J1057Map}
Map of (unreduced) $\chi^{2}$ for J1057$-$5226 in the $\alpha$--$\zeta$ plane with $3\sigma$ contours for sets of $\rho_{\rm{max}}$ ranges, 
the single altitude model fit,
and $\Delta\rho>0.2R_{\rm{LC}}$.  The best fit for the RVM is also indicated on the map.  The best fit using the RVM
is quite different from the best fit using finite altitude and restrictions from the phase of emission
seen in the data ($\Delta\rho>0.2R_{\rm{LC}}$).
}
\end{center}
\vskip -0.2truecm
\end{figure}

J1057$-$5226 is a relatively young pulsar ($P=197.11$ ms) that has had its radio polarization
position angle sweep fit in the literature with the RVM \citep{weltevrede2009mapping}.  
Here we fit the latest polarization data for J1057$-$5226 at 1.369 GHz with 
the RVM plus our own finite altitude model.  Polarization and intensity 
data for J1057$-$5226 is plotted in 
Figure~\ref{fig:PlotJ1057intPA}.  First note that we are unable
to explain the polarization position angles associated
with $C_{3}$ as labeled in Figure~\ref{fig:PlotJ1057intPA}
with either the RVM or our current model.
This portion of the polarization sweep had not appeared in the previous 
RVM fitting papers due to poor signal-to-noise.  Although we will not
attempt to explain this component here, we hope to explore modifications to 
our current model that will explain this component in future work.

Ignoring the polarization position angles associated
with $C_{3}$ for the moment, unlike J0023$+$0923 and J1024$-$0719,
J1057$-$5226 does not have any compelling features to indicate mode jumps or
multiple altitudes.  The $\chi^2_{\rm{min}}$ for the RVM even gives a reasonable fit for the 
number of degrees of freedom (DOF, Table~\ref{tb:fitJ1024}).  Even so, 
by fitting to a finite altitude and 
finite multiple altitudes, we can significantly decrease the 
$\chi^2_{\rm{min}}$ and open the parameter space.  Further, \citet{weltevrede2009mapping}
were forced to conclude that emission comes from outside the
formal open zone cap due to the large emission phase of J1057$-$5226.
Here we seek a more physical model using a finite altitude and $\rho_{\rm{ypt}}<R_{\rm{LC}}$.

We can model the data with $\rho_{\rm{ypt}}=1R_{\rm{LC}}$ within 
$1\sigma$ of $\chi^2_{\rm{min}}$ but this 
requires pushing the model to the highest allowed altitudes.  
On the other hand, low altitudes
are also permitted but require low $\rho_{\rm{ypt}}$ in 
order for the models to emit in the phase observed
in the data.  The true fit likely lies somewhere between the extremes.  A strong correlation 
exists between $\alpha$, $\zeta$, $R_{1}$, and $R_{2}$ such 
that if one has an estimated range for $R_{1}$ or $R_{2}$, the acceptable $\alpha$--$\zeta$ 
models would be significantly decreased. 

Red, cyan, and black contours on Figure~\ref{fig:J1057Map} are $3\sigma$ contours
from $\chi^2_{\rm{min}}$ of different ranges of $\rho_{\rm{max}}$, the maximum cylindrical 
distance of the emission within the emission phase for the model.  These contours exemplify
the correlation between altitude and $\alpha$--$\zeta$ pairs in a single parameter.  Further, fits with smaller
$\rho_{\rm{max}}$ are more physically likely but so is larger $\Delta \rho$ such as 
$\Delta \rho> 0.2 R_{\rm{LC}}$ ($3\sigma$ dashed green contours
on Figure~\ref{fig:J1057Map}).  As such, one might expect the most physically plausible
fits to fall within the green and cyan contours.

The yellow contour is the $3\sigma$ contour from $\chi^2_{\rm{min}}$ 
of a single-altitude model.  
Going from one altitude to two altitudes does not greatly improve $\chi^2_{\rm{min}}$.
Adding the additional parameter does allow for a wider variety of geometric
configurations ($\alpha$ and $\zeta$) which could be of importance when comparing
to multi-wavelength results.

For the RVM (unreduced) $\chi^2_{\rm{min}}=325$ and for the 
two-altitude model (unreduced) $\chi^2_{\rm{min}}=281$
(see Table~\ref{tb:fitJ1057}).
The $F$-test between the RVM and the two-altitude 
model gives $F=12.50$, DOF$_1=2$, and DOF$_2=160$. The probability of
exceeding this $F$ is $P=8.78\e{-6}$. Comparing the RVM 
to the single-altitude model ($\chi^2_{\rm{min}}=282$), the
probability of exceeding $F=24.55$ is $\rm{Prob}=1.82\e{-6}$.  
Comparing the single-altitude model to the two-altitude model, the
probability of exceeding $F=0.57$ is $\rm{Prob}=4.52\e{-1}$.

\subsection{J1744$-$1134: Multiple Altitudes with Single Versus Double Pole}
\label{sec:J1744}

\begin{table*}[t!!]
\small
\tabcolsep=0.1cm
\caption{Fit Parameters for J1744$-$1134}
\begin{center}
\begin{tabular}{lcccccccc}
\hline
\hline
&&(Unreduced)&&&&&\\
& DOF&$\chi^2_{\rm{min}}$ & $\alpha$ ($^\circ$) & $\zeta$ ($^\circ$) & $R_1$ ($R_{\rm{LC}}$)& $R_2$ ($R_{\rm{LC}}$)& $\Delta \rho$ ($R_{\rm{LC}}$)& $\rho_{\rm{ypt}}$ ($R_{\rm{LC}}$)\\ 
[.3em]\hline

RVM & 209-4&$   355 $ & $ 74 ^{+0(+1)}_{-2(-4)} $ & $97 ^{+0(+1)}_{-2(-3)} $ &  \nodata&\nodata &\nodata&\nodata \\ \hline

2 Alt no jump & 209-6&$   311 $ & $   76 ^{+2(+6)}_{-11(-15)} $ & $   60 ^{8(+84)}_{-18(-26)} $ 
& $ 0.68 ^{+0.09(+0.18)}_{-0.12(-0.62)} $ & $ 0.70 ^{+0.20(+0.20)}_{-0.03(-0.64)} $ 
& $ 0.06 ^{+0.00(+0.33)}_{-0.04(-0.06)} $ & $ 0.73 ^{+0.19(+0.27)}_{-0.03(-0.06)} $ \\ \hline

2 Alt jump & 209-6&$   314 $ & $   92 ^{+2(+4)}_{-53(-70)} $ & $57 ^{+76(+79)}_{-8(-32)} $ 
& $ 0.87 ^{+0.03(+0.03)}_{-0.76(-0.81)} $ & $ 0.88 ^{+0.02(+0.02)}_{-0.24(-0.44)} $ 
& $ 0.05 ^{+0.24(+0.27)}_{-0.05(-0.05)} $ & $ 0.81 ^{+0.19(+0.19)}_{-0.11(-0.37)} $ \\ \hline

1 Alt single pole & 209-5& $310$ & $66 ^{+5(+15)}_{-4(-7)}$ & $85 ^{+3(+35)}_{-36(-39)}$  
& $0.65 ^{+0.07(+0.20)}_{-0.27(-0.54)} $ & \nodata
& $ 0.36 ^{+0.09(+0.12)}_{-0.35(-0.47)} $ & $ 0.99 ^{+0.01(+0.01)}_{-0.50(-0.86)} $ \\ \hline

2 Alt single pole & 209-6& $309 $ & $66 ^{+11(+17)}_{-22(-36)} $ & $85 ^{+24(+32)}_{-53(-60)} $ 
& $ 0.65 ^{+0.14(+0.25)}_{-0.55(-0.59)} $ & $ 0.59 ^{+0.31(+0.31)}_{-0.23(-0.52)} $ 
& $ 0.37 ^{+0.19(+0.30)}_{-0.37(-0.37)} $ & $ 1.00 ^{0.00(+0.00)}_{-0.56(-0.93)} $ \\ \hline

\end{tabular} 
\tablecomments{Errors reported without (with) parentheses are for $1\sigma$ ($3\sigma$) from $\chi^2_{\rm{min}}$. }
\label{tb:fitJ1744-1134} 
\end{center} 
\end{table*}

\begin{figure*}[t!!]
\vskip .4\textheight
\includegraphics{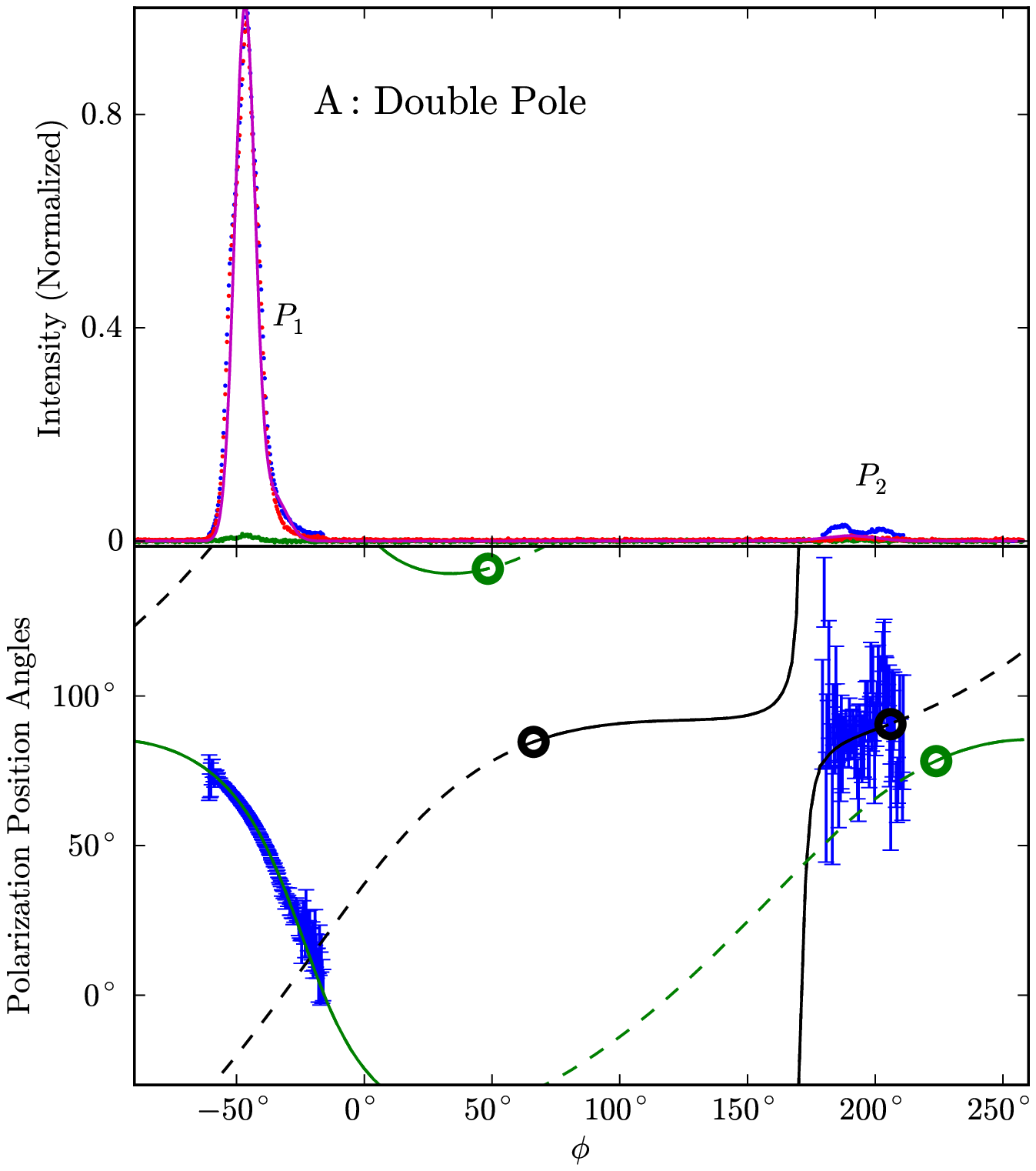}
\includegraphics{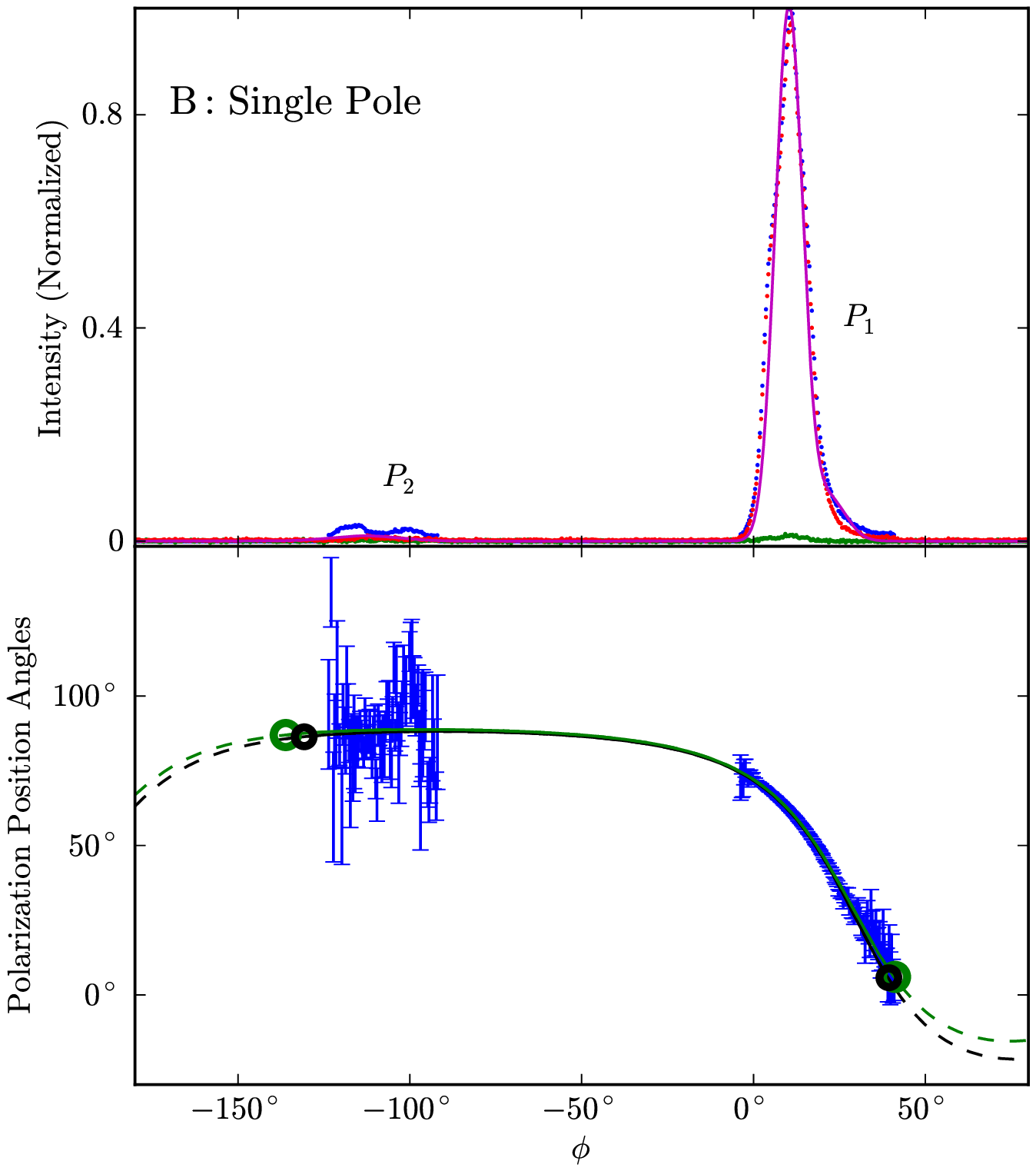}
\begin{center}
\caption{\label{fig:PlotJ1744intPA}
In the upper panels, blue points are total radio intensity data for 1.369 GHz, red points are linear polarization intensity data,
and green points are circular polarization intensity data for J1744$-$1134.  The solid magenta line in the upper panels is
the model linear intensity used in fitting.  In the bottom panels, blue error bars are polarization position angles
used in the fit.
For panel (A), the model polarization comes from a double magnetic pole model with (unreduced) $\chi^2=342$, $\alpha=82^{\circ}$, and
$\zeta=39^{\circ}$.  The green line is the polarization for
a model with $R_{1}=0.78R_{\rm{LC}}$ and the black line is the polarization for 
a model with $R_{2}=0.72R_{\rm{LC}}$.
The emission phase from the data for these model parameters 
requires $\rho_{\rm{ypt}}=0.96R_{\rm{LC}}$ as marked on the plot with the solid lines.
For panel (B), the model polarization comes from a single magnetic pole model with (unreduced) $\chi^2=317$, $\alpha=66^{\circ}$, and 
$\zeta=85^{\circ}$.  The green solid line is the polarization for
a model with $R_{1}=0.65R_{\rm{LC}}$ and the black solid line is the polarization for 
a model with $R_{2}=0.59R_{\rm{LC}}$.  
The emission phase from the data is covered with $\rho_{\rm{ypt}}=1R_{\rm{LC}}$ for these model parameters.
Empty circles mark the limiting phase of emission from
open field lines with $\rho_{\rm{ypt}}=1R_{\rm{LC}}$.
A phase of zero is the point
of closest encounter to the magnetic axis in the model.
}
\end{center}
\vskip -0.2truecm
\end{figure*}

\begin{figure}[t!!]
\vskip .38\textheight
\includegraphics{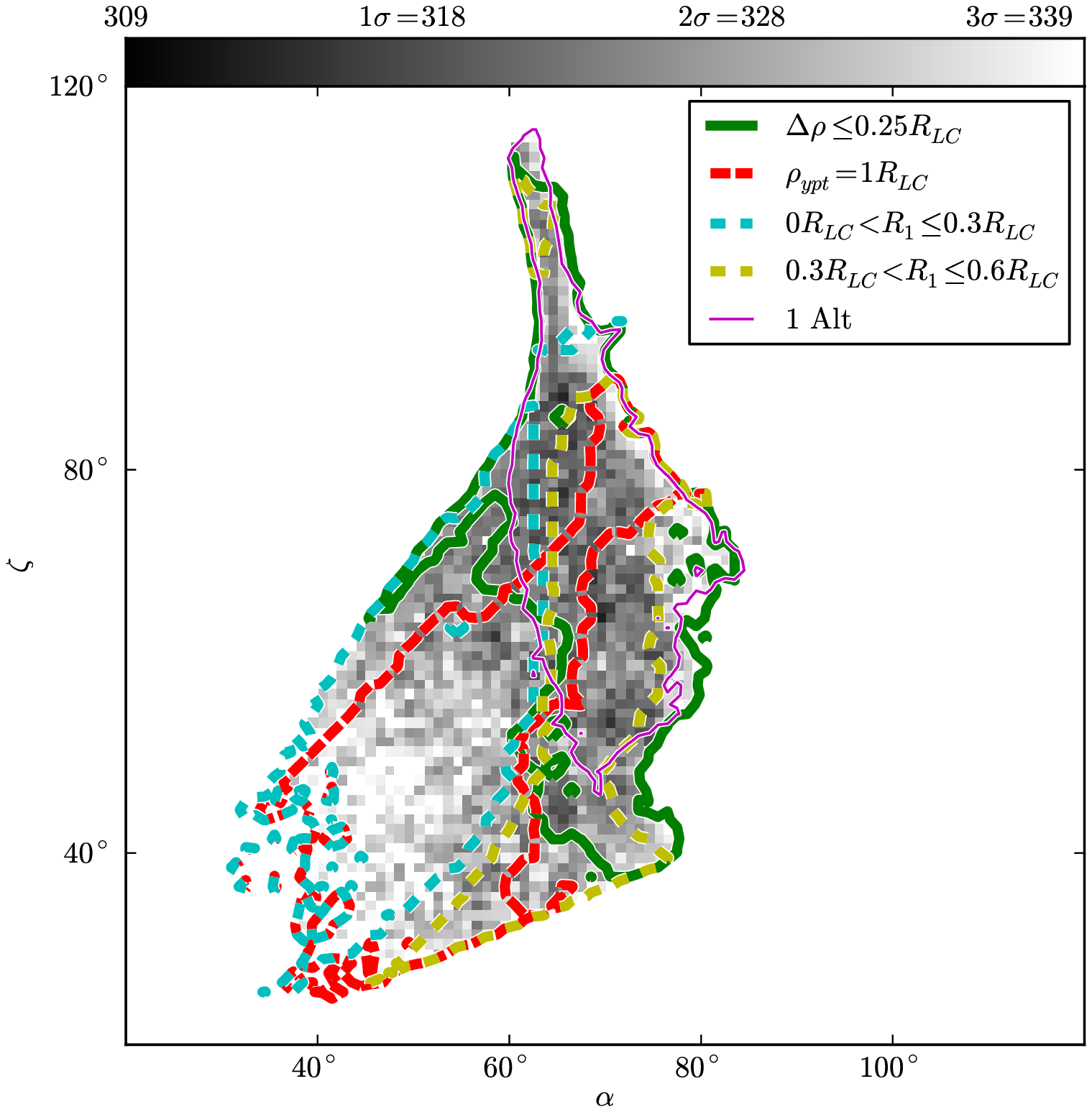}
\begin{center}
\caption{\label{fig:J1744-1134MapNoJumpSP}
Map of (unreduced) $\chi^{2}$ for J1744$-$1134 in the $\alpha$--$\zeta$ plane for a single magnetic pole model with $3\sigma$ 
contours for two $R_{1}$ ranges to show correlation between $\alpha$, $\zeta$, and altitude for
$\Delta\rho\leq 0.25R_{\rm{LC}}$ (the more approximate models) and $\rho_{\rm{ypt}}=1R_{\rm{LC}}$
(the less approximate models), and for the single-altitude model which lies mostly 
in the physically inaccurate contour of $\Delta\rho\leq 0.25R_{\rm{LC}}$ and therefore argues for a 
two-altitude model.
}
\end{center}
\vskip -0.2truecm
\end{figure}

J1744$-$1134 is yet another millisecond pulsar detected in $\gamma$-rays 
by \textit{Fermi}.  Similar to J1057$-$5226, 
the polarization of J1744$-$1134 fits well to the RVM.  
But even with the consideration that the star surface is 
$\sim 0.06R_{\rm{LC}}$ based on period, the emission zone
of a vacuum dipole model with emission from both poles is not large enough to accommodate 
the range of the emission in phase seen in the data. 
In the vast majority of fits, the cylindrical distance between
the edge of the open zone required and the maximum cylindrical emission
point ($\Delta \rho$) is smaller than $0.2R_{\rm{LC}}$.  We considered a
two-pole model with and without an orthogonal mode jump between the polarization
position angles associated with $P_1$ and $P_2$ as labeled on Figure~\ref{fig:PlotJ1744intPA}.
Results for the fits are reported in Table \ref{tb:fitJ1744-1134}.  

The peaks, $P_1$ and $P_2$, are separated by $\sim 103^{\circ}$--$134^{\circ}$; thus another possibility is that the emission
is not from two magnetic poles but a single broad pulse.  Interestingly, with this assumption,
models exist within the $1\sigma$ multidimensional contour with $\rho_{\rm{ypt}}=1R_{\rm{LC}}$.
Figure~\ref{fig:J1744-1134MapNoJumpSP} shows the $\alpha$--$\zeta$ maps of (unreduced) $\chi^2$
for the single broad pulse model.  The red contour shows the $3\sigma$ range with the 
assumption that $\rho_{\rm{ypt}}=1R_{\rm{LC}}$. Additionally, the green contour is the $3\sigma$
contour with a $\Delta \rho \leq 0.25R_{\rm{LC}}$ cut and the thin magenta line is the $3\sigma$
contour if $R_1=R_2$; these two contours strongly overlap which makes us slightly favor models
where $R_1\neq R_2$.  Quite a range of altitudes falls 
within the $3\sigma$ contours and there is
a strong correlation between $R_1$ and $\alpha$ and $\zeta$.  We also plotted two rough ranges
of $R_1$ on the $\alpha$--$\zeta$ map to illustrate this correlation.
Overall, the polarization position angles associated 
with $P_2$ are very noisy, which translates
into noisy $\chi^2$ surfaces.  To decrease this noise, 
we applied a $0\fdg5$ Gaussian smoothing kernel
to the map and contours.  Additionally, Figure~\ref{fig:PlotJ1744intPA},
panel (B) shows the polarization sweep derived from a
single magnetic pole model with $\alpha=66^\circ$ and $\zeta=85^\circ$.

For single-pole and double-pole models, fit parameters and errors are reported in Table~\ref{tb:fitJ1744-1134}.
Overall, adding a finite altitude, whether using 
a single-pole or two-pole model, significantly decreases the $\chi^2_{\rm{min}}$
which can be shown statistically using an $F$-test.  
The (unreduced) $\chi^2_{\rm{min}}=355$ for the RVM and $\rm{DOF}_1=2$, and $\rm{DOF}_2=203$
for the $F$-test.  For the single magnetic pole emission 
(unreduced $\chi^2_{\rm{min}}=309$), the probability of exceeding the resulting $F$ is $P=7.63\e{-7}$;
for the two magnetic pole model without 
an orthogonal mode jump (unreduced $\chi^2_{\rm{min}}=311$), 
the probability of exceeding the resulting $F$ is 
$P=1.47\e{-6}$; for the two magnetic pole model 
with an orthogonal mode jump (unreduced $\chi^2_{\rm{min}}=314$), the probability of 
exceeding the resulting $F$ is $P=3.89\e{-6}$.

\subsection{J1420$-$6048: Multiple Altitudes and Interstellar Scattering}
\label{sec:J1420}

\begin{figure*}[t!!]
\vskip .38\textheight
\includegraphics{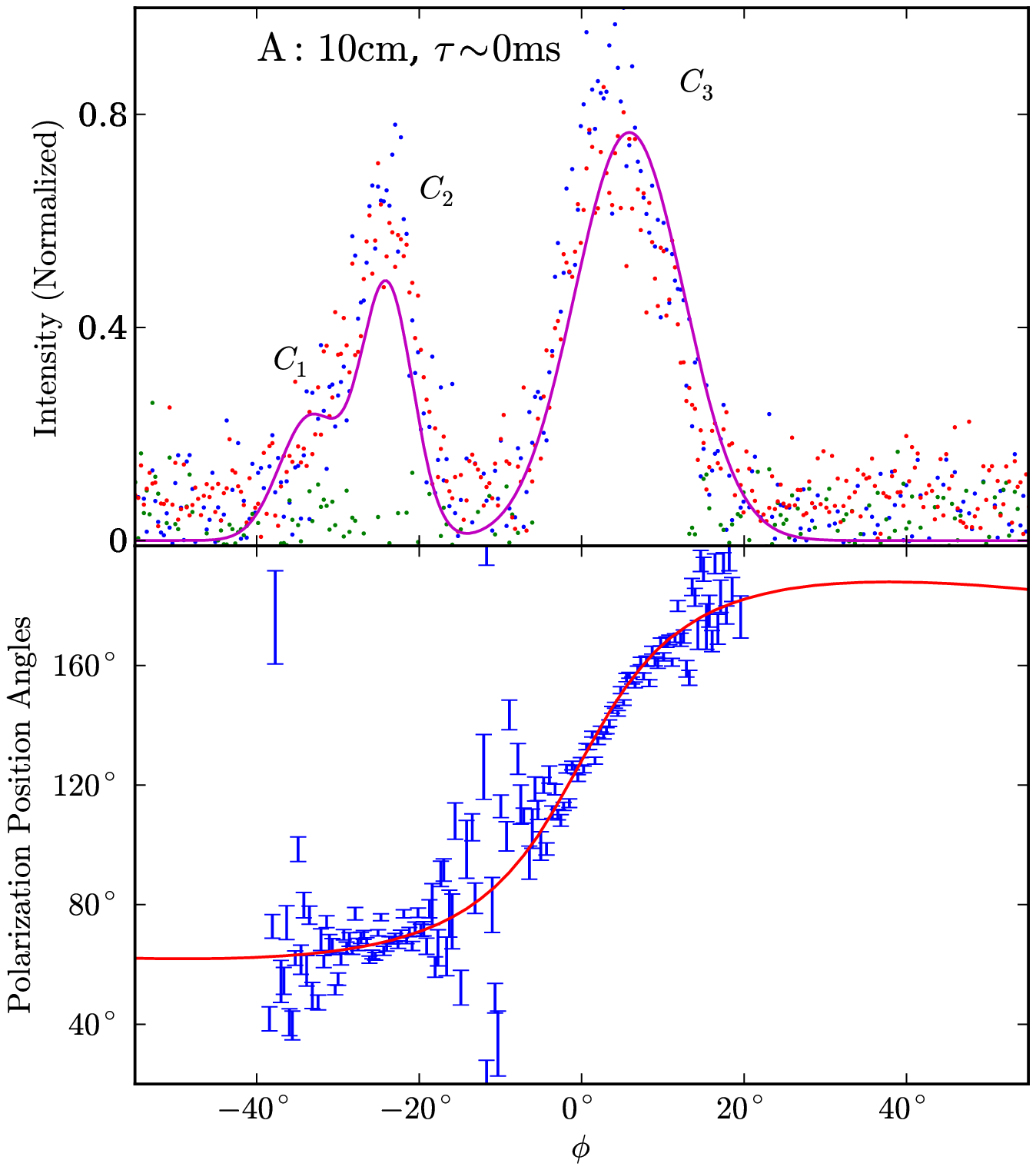}
\includegraphics{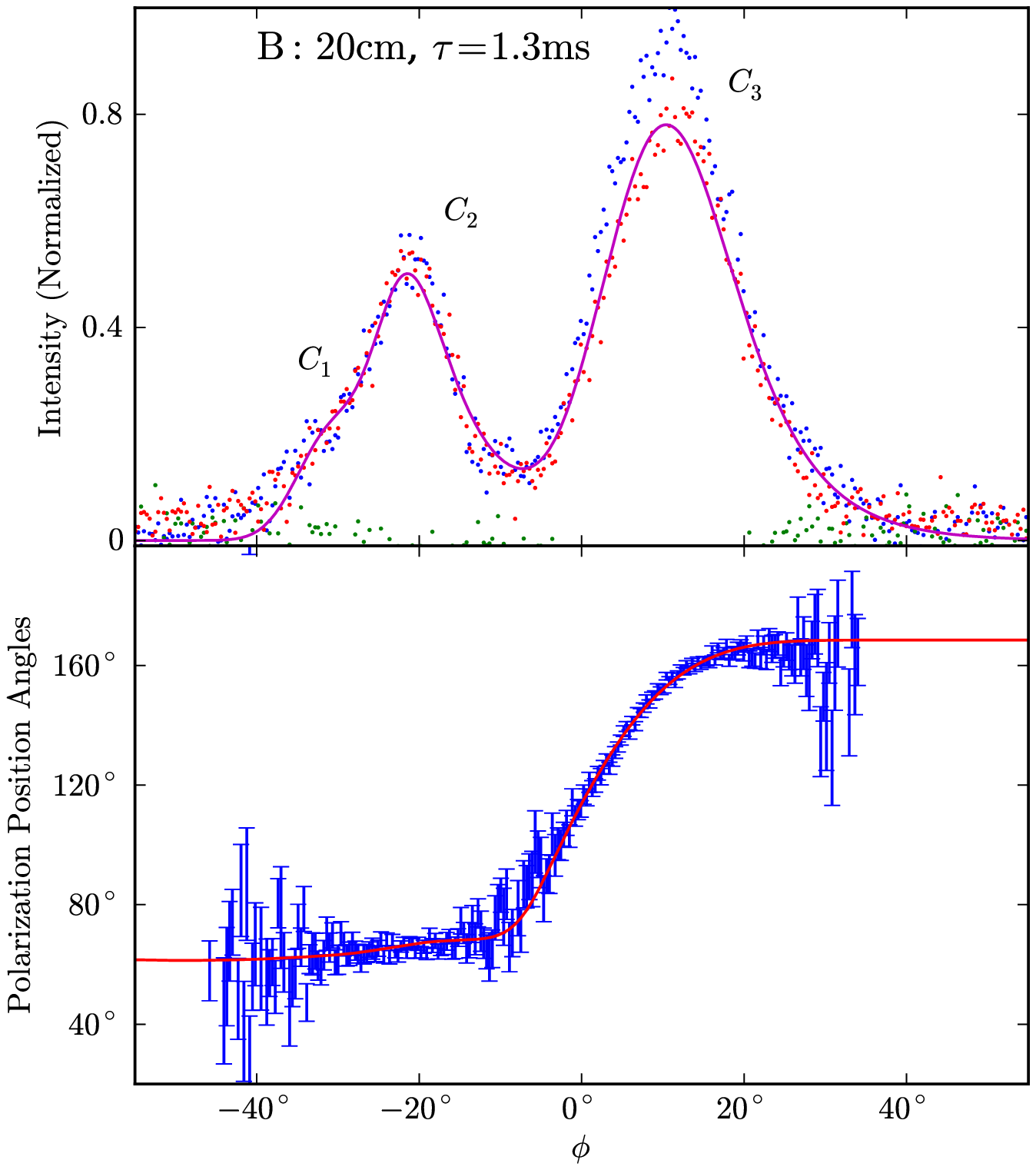}
\begin{center}
\caption{\label{fig:PlotJ1420intPA}
In the upper panel, blue points are total radio intensity data, red points are linear polarization intensity data,
and green points are circular polarization intensity data for J1420$-$6048.  The solid magenta line in the upper panel is
the model linear intensity used in fitting.  In the bottom panel, blue error bars are polarization position angles
used in the fit.
The model polarization comes from
a fit with (unreduced) $\chi^2=435$ (joint fit with data from both 10 cm and 20 cm) 
and parameters $\alpha=120^{\circ}$, $\zeta=150^{\circ}$, and $R_{1}=0.53R_{\rm{LC}}$ weighted by the model intensity.
Panel (A) shows the 10 cm intensity and polarization position angle data and the model with scattering time $\tau\sim0$ ms.
Panel (B) is the 20 cm intensity and polarization position angle data and the model with scattering time $\tau=1.3$ ms.
The $\rho_{\rm{ypt}}<1R_{\rm{LC}}$ constraints lie beyond the phase plotted.
A phase of zero is the point
of closest encounter to the magnetic axis in the model.
}
\end{center}
\vskip -0.2truecm
\end{figure*}

\begin{figure}[t!!]
\vskip .35\textheight
\includegraphics{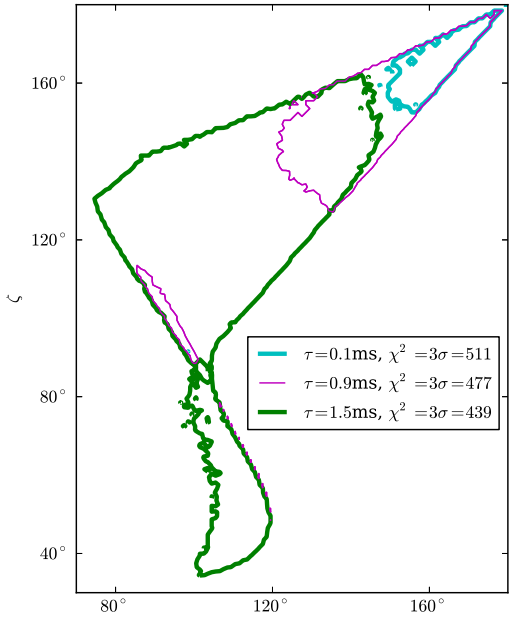}
\begin{center}
\caption{\label{fig:diffTau}
Contours of the joint (unreduced) $\chi^2$ set to $3\sigma$ in the $\alpha$--$\zeta$ plane for modeling of 
the polarization data in 10 cm and 20 cm of J1420$-$6048.  
Different colors represent fits to different $\tau$, scattering time.  
Increasing scattering time widens the acceptable fit parameters, decreases the acceptable $\alpha$ 
values, and decreases $\chi^2_{\rm{min}}$.
}
\end{center}
\vskip 0.3truecm
\end{figure}

\begin{figure}[t!!]
\vskip .36\textheight
\includegraphics{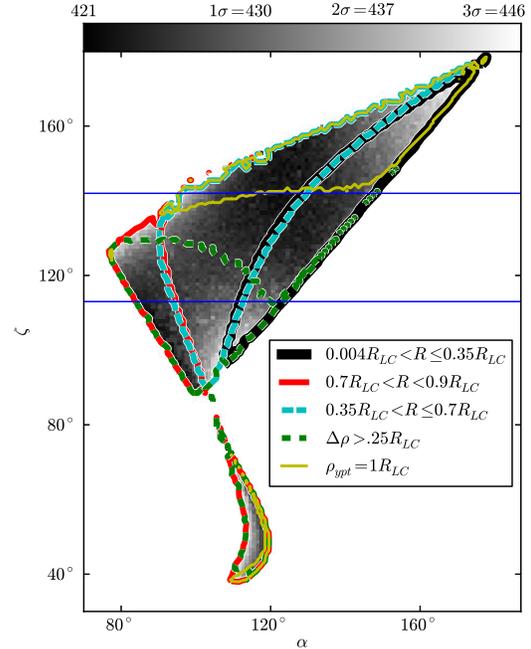}
\begin{center}
\caption{\label{fig:T1.3map}
Map of the joint (unreduced) $\chi^{2}$ for polarization data in 10 cm and 20 cm of J1420$-$6048 in the $\alpha$--$\zeta$ plane. 
The model has a scattering constant of $\tau=1.3$ ms.
Contours of $3\sigma$ are for three ranges of $R$ to show the correlation between $\alpha$, $\zeta$, and altitude and for
$\Delta\rho\leq 0.25R_{\rm{LC}}$ (the most physically inaccurate models) and $\rho_{\rm{ypt}}=1R_{\rm{LC}}$
(the most physically accurate models).  Horizontal blue lines indicate the region favored by X-ray
torus fitting (Section \ref{sec:xray}).
}
\end{center}
\vskip -0.15truecm
\end{figure}

The millisecond pulsar ($P=68$ ms) J1420$-$6048 has 
been studied previously (\citealt{roberts2001multiwavelength};
\citealt{weltevrede2010gamma}).  
The RVM fit in these papers estimated $\alpha>145^\circ$ and $\zeta-\alpha\sim 0\fdg5$
(the angle convention used in these papers is different from that used
in this paper and they must be converted.  
See \citealt{everett2001emission} for an explanation and the conversion formula).
This fit is consistent with our results of fitting with the RVM.
Since the effects of interstellar scattering scales 
to the $-4$ power in frequency \citep{lang1971pulse}, 
comparing 10 cm data to 20 cm data reveals that 
the polarization and intensity profile of J1420$-$6048 have 
signs of scattering (Figure~\ref{fig:PlotJ1420intPA}).  
In particular, note the widening of the intensity profile and the flattening
of the polarization sweep at the trailing edge of the intensity components for 
the 20 cm data compared to the 10 cm data.  The scattering
time constant calculated using the \citet{cordes2002ne2001} 
model is $\tau=6.5\e{-2}$ ms.  The ratio of $\tau$
to period (and also the value of dispersion measure or DM) is far larger than that for any of the other
pulsars considered in this paper yet this value is still smaller than 
what would cause the scattering seen by comparing the 10 cm and 20 cm data.
Because of this discrepancy, $\tau$ is a fit parameter in analyzing 
polarization data of J1420$-$6048. 

When fitting the 10 cm data ($\tau\sim0$), the best fit non-scattered model 
is $\alpha=100^{+34(+79)\circ}_{-63(-97)}$, $\zeta=39^{+116(+140)\circ}_{-25(-38)}$, and $R=0.64^{+0.26(+0.26)}_{-0.64(-0.64)}R_{\rm{LC}}$
where errors without (with) parentheses are $1\sigma$ ($3\sigma$) errors.
The error bars on these values are large due to 
the low signal-to-noise in the polarization position
angles and the exact parameter values at 
$\chi^2_{\rm{min}}$ are less valuable than the full range defined by these
error bars.  When fitting the 20 cm polarization data, 
with a small scattering time ($\tau=0.1$ ms), 
the best fit non-scattered model is $\alpha=175^{+2(+3)\circ}_{-6(-18)}$, $\zeta=177^{+1(+1)\circ}_{-4(-14)}$, and 
$R=.37^{+0.10(+0.50)}_{-0.33(-0.37)}R_{\rm{LC}}$ which is drastically different
than the results for the 10 cm data.  In fact the $3\sigma$ multi-dimensional contours as
measured from the individual $\chi^2_{\rm{min}}$ do not overlap.

Physically, we expect that two different 
frequencies should come from different altitudes but we
are making the assumption that they are closely spaced and any systematic error from this assumption
is overpowered by the statistic error.
A combined $\chi^{2}$ from the two sets of data 
(10 cm and 20 cm) at low scattering constants results in 
a minimum region similar to the minimum of fitting 20 cm 
alone (see Figure~\ref{fig:diffTau}, cyan contours $\tau=0.1$ ms).
A combined $\chi^{2}$ from the two sets of data at 
an intermediate scattering constant results in
two distinct minimum regions (see Figure~\ref{fig:diffTau}, 
magenta contours $\tau=0.9$ ms).  
A combined $\chi^{2}$ from the two sets of data 
at high scattering constants results in
a single minimum region (see Figure~\ref{fig:diffTau}, 
green contours $\tau=1.5$ ms).
In the combined $\chi^{2}$, we assume that the altitudes are the same.
Fitting the data with different scattering constants gives drastically different results.
The larger the scattering constant, the better the $\chi^{2}_{\rm{min}}$.  
We can place a practical
limit on the scattering because a large scattering constant results in a distorted intensity 
profile which is not seen in the data.  
Therefore, we did not fit with a scattering constant larger
than $\tau=1.5$ ms since scattering constants much 
larger than this distort the intensity profile.
In Figure~\ref{fig:diffTau}, as the scattering constant increases, the two islands of 
best fit $\chi^2$ seen at the lowest scattering constant merge into a single $\chi^2$ surface.
Because of the drastic decrease in $\chi^{2}_{\rm{min}}$ from increasing the scattering
constant and the merging of the $\chi^{2}_{\rm{min}}$, the true scattering constant 
is $\tau\sim 1.1$--$1.5$ ms.  Additionally, we report $\alpha$, $\zeta$, and $R$ in
Table~\ref{tb:fitJ1420} for select values of $\tau$. 

Further, the error bars on the DM are larger than the shift that we 
expect from including scattering when comparing the two different wavelengths.
The DM reported in \citet{weltevrede2010gamma} 
is $360^{+2}_{-2}$ cm$^{-3}$ pc and the correction to the 
DM from our fitting of scattering time constants with 
error bars are reported in Table~\ref{tb:fitJ1420}.
The $\Delta \rm{DM}$ values are all with in the 2 cm$^{-3}$ 
pc error bars of the original DM up to 
$3\sigma$ from $\chi^2_{\rm{min}}$.

As $\tau$ increases, the best fit values of $\alpha$ and $\zeta$ shift and the $3\sigma$ range
for these values increases drastically.  
Also, the $\chi^2_{\rm{min}}$ values decrease statistically 
significantly from $\tau=0.1$ ms to $\tau=1.5$ ms.  
For $\tau=1.5$ ms, $3\sigma=439$ from $\chi^2_{\rm{min}}$.
The $\chi^2_{\rm{min}}$ for $\tau=0.1$ ms is not within 
this $3\sigma$ of the $\tau=1.5$ ms fit.

Figure~\ref{fig:T1.3map} is the (unreduced) $\chi^2$ map for $\tau=1.3$ ms.  The black,
cyan, and red contours are for $3\sigma$ contours at various altitude ranges, 
illustrating that although the allowed range of altitudes is large for this pulsar, knowledge
of $\alpha$ and $\zeta$ could greatly decrease this 
range because of the correlation between $R$ and
$\alpha$--$\zeta$ pairs.  A large number of fits could be additionally excluded if
cuts of $\Delta\rho$ are applied. 
The green dashed contour corresponds to $\Delta\rho<0.25R_{\rm{LC}}$.
If only fits up to $3\sigma$ with $\rho_{\rm{ypt}}=1R_{\rm{LC}}$ are considered, 
only fits within the yellow
contour on Figure~\ref{fig:T1.3map} would be allowed.

\begin{table}[ht]
\footnotesize
\tabcolsep=0.05cm
\caption{Fit Parameters for J1420$-$6048}
\begin{center}
\begin{tabular}{lccccccccc}
\hline
\hline
 &$\tau$&(Unreduced)&&&\\
DOF&(ms) & $\chi^2_{\rm{min}}$ & $\alpha$ ($^\circ$) & $\zeta$ ($^\circ$) & $R$ ($R_{\rm{LC}}$)\\ 
[.3em]\hline
356-5&.1 & $   480 $ & $  175 ^{+3(+4)}_{-15(-77)} $ & $  177 ^{+1(+2)}_{-13(-86)} $  & $ 0.37 ^{+0.10(+0.52)}_{-0.31(-0.37)} $ \\ \hline
356-5&.9 & $   448 $ & $  166 ^{+13(+13)}_{-76(-81)} $ & $  169 ^{+10(+10)}_{-104(-119)} $  & $ 0.26 ^{+0.64(+0.64)}_{-0.26(-0.26)} $ \\ \hline
356-5&1.3 & $   421 $ & $  126 ^{+22(+51)}_{-46(-50)} $ & $  153 ^{+9(+25)}_{-113(-115)} $  & $ 0.52 ^{0.38(+0.38)}_{-0.52(-0.52)} $ \\ \hline
356-5&1.5 & $   415 $ & $  105 ^{+27(+42)}_{-28(-31)} $ & $  140 ^{+14(+22)}_{-102(-106)} $  & $ 0.58 ^{+0.32(+0.32)}_{-0.44(-0.58)} $ \\ \hline

\end{tabular}
\tablecomments{Errors reported without (with) parentheses are for $1\sigma$ ($3\sigma$) from $\chi^2_{\rm{min}}$. }
\label{tb:fitJ1420}
\end{center}
\end{table}

\vskip 1.0truecm
\subsubsection{X-Ray Torus of J1420$-$6048}
\label{sec:xray}
\begin{figure}[t!!]
\vskip .27\textheight
\includegraphics{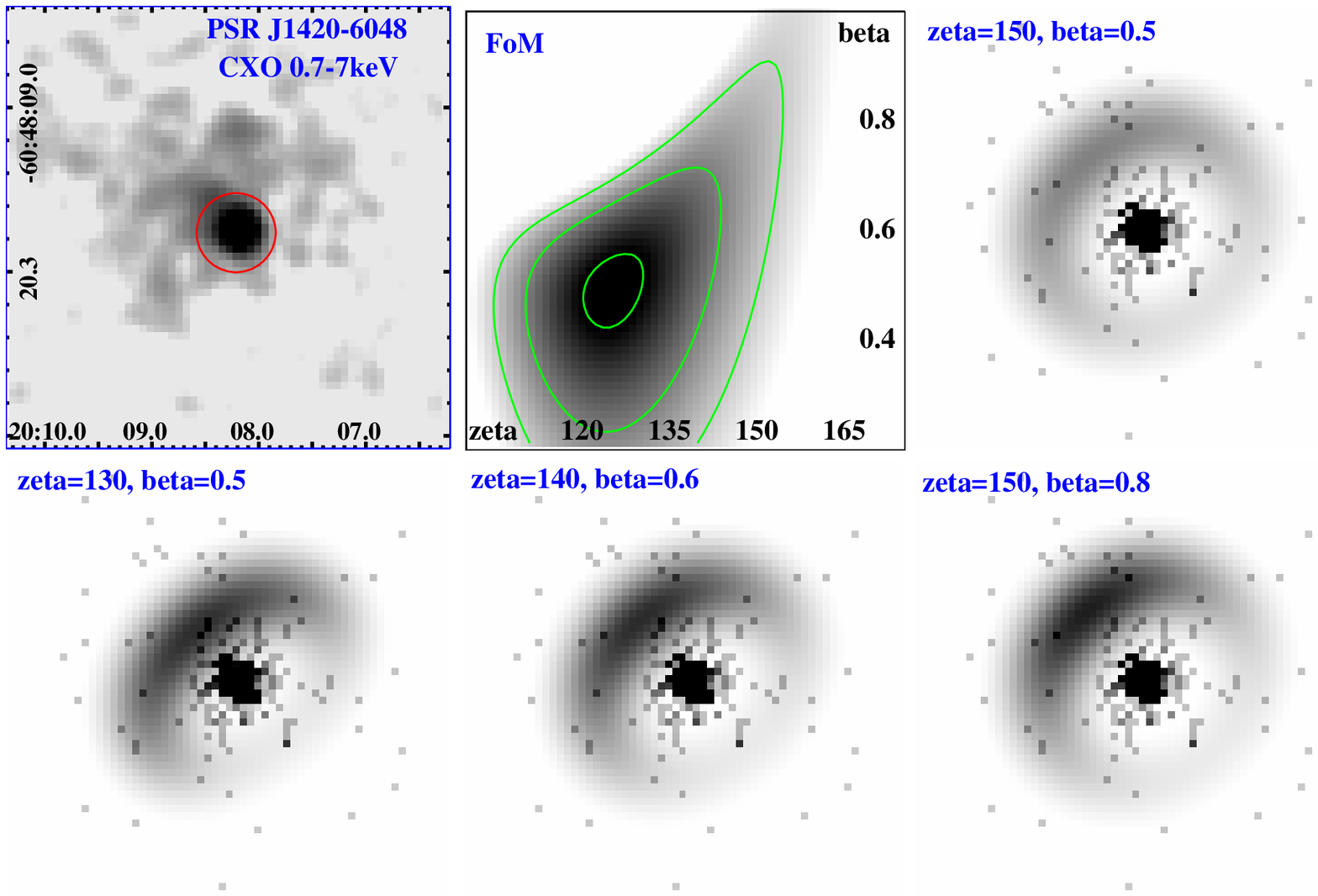}
\begin{center}
\caption{\label{fig:xraytorus}
X-ray PWN of PSR J1420$-$6048. Upper left shows the PWN structure,
while the figure of merit (FoM) panel show contours of the fit in the $\zeta$--$\beta$ plane. The
other panels show the dependence of the torus shape and brightness on these 
parameters (see Section \ref{sec:xray}).
}
\end{center}
\vskip -0.15truecm
\end{figure}

Young pulsars appear to produce relativistic plasma confined
to the spin equator. When this wind shocks the $e^+/e^-$ pitch angle scatter and
synchrotron radiate, producing an equatorial torus. Since the bulk flow in the
region is mildly relativistic, with expected bulk velocity $\beta \sim 0.3$--$0.7$,
this torus can be Doppler brightened on the side emerging from the plane of the sky.
In several cases, there seems to be a secondary shock along the polar axis producing
polar ``jets'' in the pulsar wind nebula (PWN), 
with the jet on the opposite side of the Doppler-enhanced torus
rim, being Doppler boosted. The Crab and Vela pulsars provide classic examples of this
relativistic torus--jet geometry.

\citet{ng2004fitting} and \citet{ng2008fitting} 
showed how fits to {\it Chandra} X-ray images of such tori can provide useful constraints
on the pulsar spin orientation. For PSR J1420$-$6048, we obtained a 90\,ks {\it Chandra}
ACIS observation (ObSID 12545). We combined this exposure with a 10\,ks archival observation
(ObsID 2794), removing the spacecraft dither and reprocessing the data giving
sub-pixel (EDSER) event positioning, to obtain the best possible image of the compact
PWN surrounding this energetic pulsar.  Figure ~\ref{fig:xraytorus} (upper left) shows a lightly smoothed
0.7--7 keV {\it Chandara} image of the combined 
exposure, with a logarithmic stretch. The pulsar point
source is in the red circle. Unfortunately, this pulsar does not show a striking torus
structure, so unlike several other young pulsars, we cannot obtain a high-quality,
model-independent measurement of its spin geometry. Still, the diffuse counts do show
a semi-circular arc of flux, trailing off to the NE. {\it If} we interpret this as an
equatorial torus, we can apply the methods of \citet{ng2008fitting} to constrain the spin
orientation. A few parameters are well measured: the position angle of the symmetry axis
($\Psi = 40^\circ \pm 3^\circ$, measured N through E) and the radius of the ``torus''
($7^{\prime\prime}\pm1 \farcs 5$) are reasonably 
constrained. In unsmoothed images there is
some evidence for a polar component on a 
$1^{\prime\prime}$--$2^{\prime\prime}$ scale, but this is not well
measured. The parameter of greatest interest to the present study is the inclination
$\zeta$ of the pulsar spin to the Earth line-of-sight.
To minimize sensitivity to point source flux and possible jet structure we fit outside
of the $5^{\prime\prime}$ radius red circle dominated by the central point source.
The main constraint comes from the shape and brightness ratio between the front and
back sides of the torus. The second panel of 
Figure ~\ref{fig:xraytorus} shows that this introduces
substantial co-variance between $\zeta$ and the bulk $\beta$ of the post-shock flow.
The best fits are near typical $\beta \sim 0.5$, with $\zeta \approx 125^\circ$.
This is in considerable tension with the larger  $\zeta$ preferred by the polarization position angle
fits to models with small scattering times; 
we only reach  $\zeta \approx 155^\circ$ with a rather aphysical $\beta \sim 0.9$.
This co-variance is visible along the bottom row of torus (+PSF+jet) models, which show that
as $\zeta$ increases from $130^\circ$ to $150^\circ$, the post-shock Lorentz factor
must grow to maintain a reasonable intensity ratio between the ``front'' and ``back''
sides of the torus. The last panel on the top row shows how with $\zeta=150^\circ$,
$\beta=0.5$, the torus is too face-on and uniform for a good fit to the data.
Thus the X-ray PWN structure agrees with the $\gamma$-ray pulse shape, where the observed
peak separation $\Delta = 0.31$ implies $\zeta \approx 110^\circ$--$140^\circ$, with the largest
$\zeta$ only available in the two-pole caustic picture \citep{romani2010constraining}, which tends to produce
too much unpulsed emission.

Polarization position angle fits with scattering constant $\tau=1.3$ ms as 
discussed in Section \ref{sec:J1420} favor $\zeta$ between $120^\circ$ and $150^\circ$
as can be seen from the color map of Figure~\ref{fig:T1.3map}.  
The measurement of $\zeta$ from the 
X-ray torus fit indicates $\zeta$ between $113^\circ$ 
and $142^\circ$ (from the contours of $2\sigma$).  
The horizontal blue lines on Figure~\ref{fig:T1.3map} 
mark the region of $2\sigma$ set by the X-ray 
torus fitting.  By assuming a
scattering constant, we not only reconcile the fits of 10 cm and 20 cm data but also 
find consistency between radio polarization position angle fits and X-ray torus fits. 

\subsection{J2124$-$3358: A Complex Example}
\label{sec:J2124}

\begin{table*}[t]
\tabcolsep=0.1cm
\caption{Fit Parameters for J2124$-$3358}
\begin{center}
\begin{tabular}{lccccccc}
\hline
\hline
& DOF&(Unreduced) $\chi^2_{\rm{min}}$ & $\alpha$ ($^\circ$) & $\zeta$ ($^\circ$) & $R_1$ ($R_{\rm{LC}}$)& $R_2$ ($R_{\rm{LC}}$)& $R_3$ ($R_{\rm{LC}}$)\\ 
[.3em]\hline 

RVM& 536-4&$  2331 $ & $    2 ^{+3(+7)}_{-0(-0)} $ & $    5 ^{+8(+19)}_{-0(-0)} $  &\nodata &\nodata &\nodata \\ \hline

3 Alt & 536-7&$   773 $ & $    2 ^{+7(+12)}_{-1(-1)} $ & $    2 ^{+7(+12)}_{-1(-1)} $ 
& $ 0.05 ^{+0.01(+0.03)}_{-0.00(-0.00)} $ & $ 0.40 ^{+0.01(+0.02)}_{-0.01(-0.03)} $ & $ 0.55 ^{+0.02(+0.04)}_{-0.01(-0.03)} $  \\ \hline

\end{tabular} 
\tablecomments{Errors reported without (with) parentheses are for $1\sigma$ ($3\sigma$) from $\chi^2_{\rm{min}}$.}
\label{tb:fitJ2124} 
\end{center} 
\end{table*}

\begin{figure}[t!!]
\vskip .4\textheight
\includegraphics{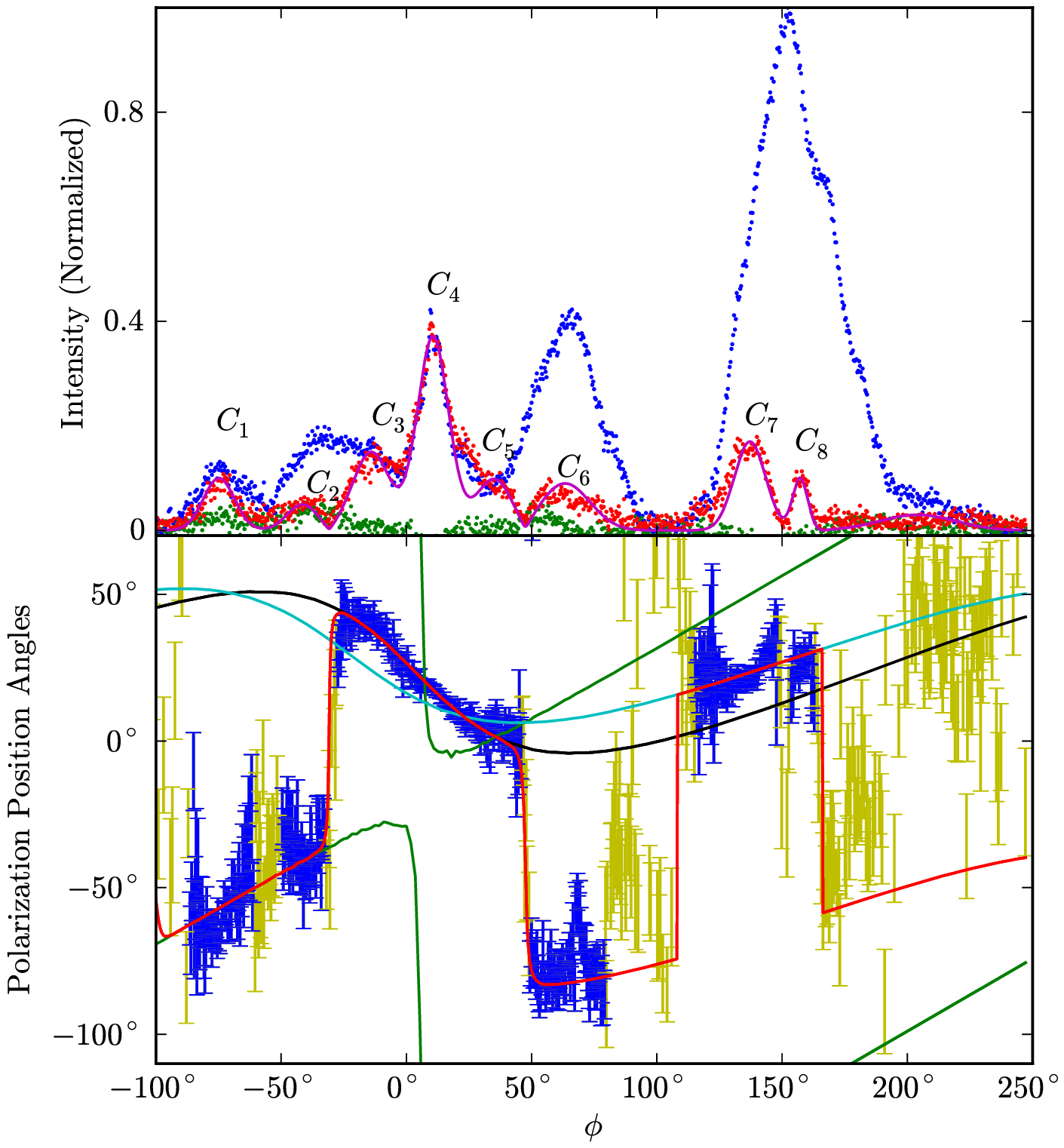}
\begin{center}
\caption{\label{fig:PlotJ2124intPA}
In the upper panel, blue points are total radio intensity data for 1.369 GHz, red points are linear polarization intensity data,
and green points are circular polarization intensity data for J2124$-$3358.  The solid magenta line in the upper panel is
the model linear intensity used in fitting.  In the bottom panel, blue error bars are polarization position angles
used in the fit and yellow error bars are polarization position angles excluded by error bar cuts.
The model polarization comes from
a fit with (unreduced) $\chi^2=773$ and parameters $\alpha=2^{\circ}$ and $\zeta=2^{\circ}$.  The green solid line is the polarization for
a model with $R_{1}=0.05R_{\rm{LC}}$ (associated with intensity components $C_1$ and $C_2$),
the black solid line is the polarization for a model with $R_{2}=0.40R_{\rm{LC}}$ (associated with intensity components $C_3$, $C_4$, and $C_5$),
and the cyan solid line is the polarization for a model with $R_{2}=0.55R_{\rm{LC}}$
(associated with intensity components $C_6$, $C_7$, and $C_8$).  
We assumed that the polarization associated with
components $C_1$, $C_2$, and $C_6$ are orthogonal to the polarization associated with
components $C_3$, $C_4$, $C_5$, $C_7$, and $C_8$.
The red solid line is the model
polarization of the three altitudes weighted by the model intensity.  There are clearly 
features in the data that are not captured by the model but the overall structure of 
the polarization is captured.
A phase of zero is the point
of closest encounter to the magnetic axis in the model.
}
\end{center}
\vskip -0.2truecm
\end{figure}

\begin{figure}[t!!]
\vskip .35\textheight
\includegraphics{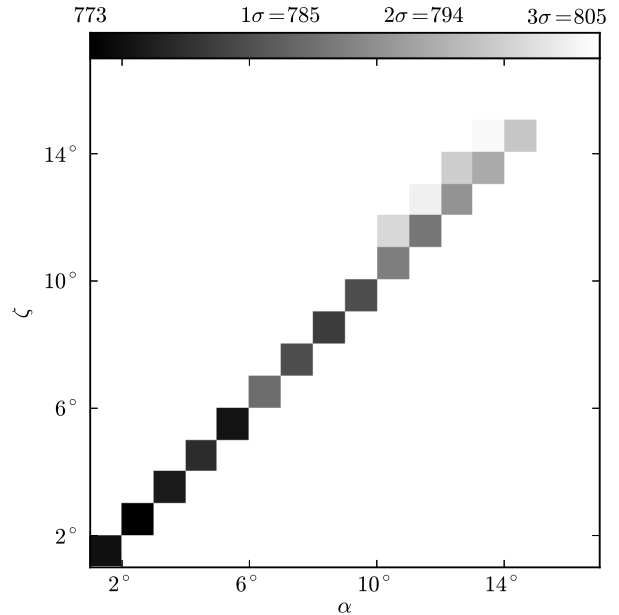}
\begin{center}
\caption{\label{fig:MapJ2124}
Map of (unreduced) $\chi^{2}$ for J2124$-$3358 in the $\alpha$--$\zeta$ plane.
All of the fitted models within $3\sigma$ from $\chi^2_{\rm{min}}$ have a phase of emission
that is within the phase of emission predicted by the models and thus $\rho_{\rm{ypt}}=1R_{\rm{LC}}$
is acceptable.
}
\end{center}
\vskip -0.2truecm
\end{figure}

J2124$-$3358 is yet another millisecond pulsar ($P=4.931$ ms).  
Plotted in Figure~\ref{fig:PlotJ2124intPA}
is the polarization and intensity profile for this pulsar at 1.369 GHz.  
The pulsar J2124$-$3358 has emission at practically 
all phases of the period.  The polarization position angles
are complicated but can be greatly simplified by assuming orthogonal mode jumps
at the appropriate components.  We assumed that the polarization associated with
components $C_1$, $C_2$, and $C_6$ are orthogonal to the polarization associated with
components $C_3$, $C_4$, $C_5$, $C_7$,  
and $C_8$ as labeled in Figure~\ref{fig:PlotJ2124intPA}.
With these orthogonal mode jumps, the polarization forms a close-to-continuous 
sweep and the RVM can be reasonably fit to the data.  Table~\ref{tb:fitJ2124} 
reports the best fit values and (unreduced) $\chi^2_{\rm{min}}$ for this fit.  

With the assumption of multiple altitudes and mode jumps, the polarization was also fit.
More than three altitudes did not significantly improve the fit.  
Also plotted in Figure~\ref{fig:PlotJ2124intPA} (top panel) 
is the best fit polarization model with multiple 
altitudes.  The polarization associated with components $C_1$ and $C_2$ is assigned one 
altitude ($R_1$); the polarization associated with components $C_3$, $C_4$, and $C_5$ is assigned the second
altitude ($R_2$); and the polarization associated with components 
$C_6$, $C_7$, and $C_8$ is assigned the third altitude ($R_3$).
The fit is far from perfect and does not capture the many bumps 
and wiggles in the polarization data.  The model does capture the overall
curvature of the polarization and significantly decreases the $\chi^2$ (Table~\ref{tb:fitJ2124})
although the $\alpha$ and $\zeta$ values do not change drastically between the two fits.
For the RVM (unreduced) $\chi^2_{\rm{min}}=2331$ and 
for the three-altitude model (unreduced) $\chi^2_{\rm{min}}=773$.
The $F$-test between RVM and the two-altitude model gives $F=355.4$, 
$\rm{DOF}_1=3$, and $\rm{DOF}_2=529$. The probability of
exceeding this $F$ is $\rm{Prob}\sim 0$, indicating the addition of altitude to the model
is highly statistically significant.

The curvature direction of the bridging polarization sweep between orthogonal
mode jumps is important here similar to the 
polarization of J0023$+$0923 in Section~\ref{sec:J0023}.
As discussed in Section~\ref{sec:MAISOM}, for a single-altitude polarization sweep
with an orthogonal mode jump, the bridging section of 
polarization between the two modes will have
the opposite curvature of that of the 
original sweep due to forward scattering.  In J2124$-$3358
between the polarization components associated with the intensity components $C_5$ and $C_6$, 
the bridging sweep direction has a negative curvature and the original sweep direction
is also negative.  This indicates that the polarization of these two components are not exactly
$90^\circ$, which is consistent with a 
multi-altitude model that has non-$90^\circ$ orthogonal jumps
between altitudes.

The values for $R_1$, $R_2$, and $R_3$ are quite restrictive and the statistical error
bars on these values are quite small.  These values are small because very
few altitude combinations capture the subtle difference between polarization
associated with the various components; for instance, note the rather large
offset between the black and cyan solid lines in Figure~\ref{fig:PlotJ2124intPA}
which represent model polarization from $R_{1}$ and $R_{2}$.  Only very 
particular sets of altitude will result in polarization with this amount
of vertical shift.
Also, $\rho_{\rm{ypt}}=1 R_{\rm{LC}}$ for all fits within
the $3\sigma$ bound of $\chi^2_{\rm{min}}$ due to the small geometrical angles 
($\alpha$ and $\zeta$, see Figure~\ref{fig:MapJ2124}).

\section{Conclusion}
\label{sec:conclusion}
In this paper, we attempted to push the limit of what we can 
learn from geometrical-based models applied to radio polarization.
We have shown that this model can explain polarization for 
which the RVM fails (partially or fully) and can significantly alter 
fit parameters ($\alpha$ and $\zeta$) obtained from the RVM.
We have shown that a handful of physical 
effects can alter our understanding of the geometry 
of millisecond and young pulsar radio emission.  
Additionally, we provided statistical comparisons to simpler 
models to quantify the significance of adding 
these physical phenomena to the model.

Both J0023$+$0923 and J1024$-$0719 clearly illustrate
how multi-altitudes can capture the non-$90^\circ$ jumps seen in the
position angle sweeps of the millisecond pulsar population.
J1057$-$5226 and J1744$-$1134 illustrate the need 
for finite altitude and $\rho_{\rm{ypt}}<R_{\rm{LC}}$
to fully explain the large phase range of the emission seen in the data.
J1420$-$6048 illustrates how scattering affects can rectify
discrepancies seen between multi-wavelength data.
Finally, J2124$-$3358 is a typical worst-case radio polarization from a millisecond pulsar.
Despite its clear non-RVM characteristics, we were
able to capture the overall structure of the polarization position angles sweep
with finite and multiple altitudes and orthogonal mode jumps.

The RVM is not accurate for the radio
polarization sweeps of these energetic pulsars.  First, this emission
originated from a significant fraction of the light cylinder which necessitates
numerical calculation of this radio polarization.  Additionally, non-$90^\circ$
jumps cannot be explained by simple orthogonal mode jumps and some polarization 
is scattered by the interstellar medium.
Polarization of millisecond pulsars is notoriously hard to model
and very few studies have attempted to tackle these objects.
That we can explain some of the polarization of these pulsars
is a significant step in the correct direction. 
This is a methods paper; with various example 
polarization data from a number of pulsars, we have shown that this 
method of using physically motivated, geometrically based 
phenomena can explain the inconsistencies of simpler models.

\acknowledgements
We greatly thank R. N. Manchester for supplying 
radio data for J1024$-$0719, J1744$-$1134, and J2124$-$3358 which is
published in \cite{yan2011polarization}.
We also owe many thanks to S. Johnston (2012, private correspondence) 
for supplying radio data for
J1057$-$5226 and J1420$-$6048 and to J. W. T. Hessels for supplying
radio data for J0023$+$0923 (J. W. T. Hessels et al. in preparation).
Roger W. Romani prepared figures and wrote the section on X-ray analysis of J1420$-$6048.
Support for this project was provided in part by grants GO1-12073X and
G03-14057A from the Smithsonian Astrophysical Observatory. This work
was also supported in part by NASA grants NNX10AP65G and NAS5-00147.
This work has been supproted by the Stanford Office of the Vice Provost
of Graduate Education DARE Doctoral Fellowship Program to H.A.C.

\bibliographystyle{apj}
\bibliography{bibtexfile}{}
\end{document}